\documentclass[lettersize,journal]{IEEEtran}
\usepackage{amsmath,amsfonts}
\usepackage{algorithmic}
\usepackage{algorithm}
\usepackage{array}
\usepackage[caption=false,font=normalsize,labelfont=sf,textfont=sf]{subfig}
\usepackage{textcomp}
\usepackage{stfloats}
\usepackage{url}
\usepackage{adjustbox}
\usepackage{verbatim}
\usepackage{graphicx}
\usepackage[table,xcdraw]{xcolor}
\usepackage{cite}
\usepackage{multirow}                 
\usepackage{multicol}                 
\usepackage{multirow}                
\usepackage{float}                    
\usepackage{makecell}                 
\usepackage{booktabs}                 
\usepackage{makecell}
\usepackage{diagbox}
\usepackage{multirow}
\usepackage {amsmath}
\usepackage {graphicx}
\usepackage {subfig}
\usepackage[switch]{lineno}
\usepackage{booktabs}
\usepackage{color}
\usepackage{enumitem}
\usepackage{graphicx}
\usepackage{lscape}
\usepackage{times}
\usepackage{soul}
\usepackage{url}
\usepackage[hidelinks]{hyperref}
\usepackage[utf8]{inputenc}
\usepackage[small]{caption}
\usepackage{graphicx}
\usepackage{amsmath}
\usepackage{float}                    
\usepackage{amsthm}
\usepackage{booktabs}
\usepackage{algorithm}
\usepackage{pifont}
\usepackage{makecell}
\usepackage{color}
\usepackage {subfig}
\usepackage{multirow}
\usepackage{algorithmic}
\usepackage{enumitem}
\usepackage{hyperref}

\usepackage{graphicx} 
\begin{document}

\title{Subjective and Objective Quality-of-Experience Evaluation Study for Live Video Streaming}

\author{Zehao Zhu, Wei Sun, Jun Jia, Wei Wu, Sibin Deng, Kai Li,  Xiongkuo Min, Jia Wang, Ying Chen, Guangtao Zhai\textit{, Fellow, IEEE}

\thanks{Corresponding author: Wei Sun, Guangtao Zhai.

Zehao Zhu, Jun Jia, Xiongkuo Min, Jia Wang, Guangtao Zhai and
Wei Sun are with the Institute of Image Communication and Network Engineering, Shanghai Jiao Tong University, Shanghai 200240, China (email: zhuzehao,jiajun0302,minxiongkuo,jiawang,zhaiguangtao@sjtu.edu.cn, sunguwei@gmail.com)

Wei Wu, Sibin Deng, Kai Li and Ying Chen are with Alibaba
Group, Hangzhou, China (email: wuwei.lorenzo@gmail.com, sibin.dsb, kaishi.lk@taobao.com, YingChen@alibaba-inc.com)
}

}



\maketitle

\begin{abstract}
In recent years, live video streaming has gained widespread popularity across various social media platforms. Quality of experience (QoE), which reflects end-users' satisfaction and overall experience, plays a critical role in media service providers to optimize large-scale live compression and transmission strategies to achieve perceptually optimal rate-distortion trade-off. Although many QoE metrics for video-on-demand (VoD) have been proposed, there remain significant challenges in developing QoE metrics for live video streaming. To bridge this gap, we conduct a comprehensive study of subjective and objective QoE evaluations for live video streaming. For the subjective QoE study, we introduce the first live video streaming QoE database, TaoLive QoE Database, which consists of $42$ source videos collected from real live broadcasts and $1,155$ corresponding distorted ones degraded due to a variety of streaming distortions, including conventional streaming distortions such as compression, stalling, as well as live streaming specific distortions like frame skipping, variable frame rate, etc. Subsequently, a human study was conducted to derive subjective QoE scores of videos in the TaoLive QoE database. For the objective QoE study, we benchmark existing QoE models on the TaoLive QoE database as well as publicly available QoE databases for VoD scenarios, highlighting that current models struggle to accurately assess video QoE, particularly for live content. Hence, we propose an end-to-end QoE evaluation model, Tao-QoE, which implicitly encodes statistical Quality of Service (QoS) features, integrates multi-scale semantic features and optical flow-based motion features to predict a retrospective QoE score. Extensive experiments demonstrate that Tao-QoE outperforms other models on the TaoLive QoE database, six publicly available QoE databases, and eight user-generated content (UGC) video quality assessment (VQA) databases, showcasing the effectiveness and feasibility of Tao-QoE.
\end{abstract}

\begin{IEEEkeywords}
quality of experience, optical flow, video quality assessment, streaming.
\end{IEEEkeywords}

\section{Introduction}


\IEEEPARstart{W}{ith} the rapid growth of mobile devices and advancements in wireless networks in recent years, people can now watch video content on mobile devices anywhere and anytime. Streaming media technologies play an important role in ensuring that users can view such content smoothly and in real-time without waiting for complete file downloads. Specifically, the streaming media content captured by the cameras or the third-party streaming media content is encoded and segmented into data fragments. These data fragments are then transmitted to the server using appropriate transport protocols. Users utilize client devices (\textit{e.g.}, mobile phones, tablets, computers, network TVs) to send requests over the Internet for accessing streaming media content. Upon receiving a client request, the server employs a content distribution network to distribute the corresponding data fragment to the requesting client device. After decoding and rendering, the data is converted into audio and video content for the user to view~\cite{ref-1}. Video on Demand (VoD) and live streaming are two prevalent methods of streaming media technology~\cite{juluri2015measurement}. Extensive research has also been conducted on quality assessment for emerging media formats such as point clouds ~\cite{zhou1}, 360-degree images ~\cite{zhou5}, omnidirectional videos ~\cite{zhou6}, light field images ~\cite{zhou7}, and stereoscopic content ~\cite{zhou8, zhou9} as well as general blind image quality assessment ~\cite{zhou2,zhou3,zhou4}, while our work specifically focuses on live streaming. In this paper, we focus on live streaming, which is the real-time broadcasting of video and audio content over the internet and allows users to watch or listen to media as it is transmitted live.


However, limited network resources and fluctuations in client networks can lead to issues such as video quality degradation, stalling events, etc., ultimately reducing QoE~\cite{ref-2,live1, seufert2014survey} of end-users. Therefore, it is essential to study how these factors influence QoE, as it can optimize more effective streaming media transmission strategy to improve users' QoE~\cite{ref-3}. Existing QoE studies rely too much on the statistical information provided by the database, including the VQA indicators of the video, the occurrence time and duration of the stalling event, etc. Many QoE models directly use these statistics as model inputs. However, the distortion in live broadcast scenarios is unpredictable. This makes the existing QoE models unsuitable for real live broadcast scenarios. At the same time, the distortion contained in the existing QoE database is more inclined to the distortion in VoD video, which is not suitable for the distortion in real live broadcast scenes.

In contrast to well-established VoD-based QoE studies, studies on live streaming QoE remain limited, primarily due to two key factors:
\begin{itemize}[leftmargin=*]
    \item \textbf{Limited live streaming QoE databases.} Existing QoE databases, such as the LIVE-NFLX series~\cite{live1,live2} and the WaterlooSQoE series~\cite{waterloo1,waterloo2,waterloo3,waterloo4}, predominantly utilize VoD setups to introduce video degradations. As a result, they fail to capture video stalling manifestations in live streaming scenarios, where network issues often result in unexpected fluctuations in frame rate and frame skipping. 
    \item \textbf{Unsatisfied QoE model mechanism.} Existing QoE models, such as KSQI~\cite{duanmu2019knowledge} and GCNN-QoE~\cite{GCNN-QoE}, rely heavily on statistical QoS features (\textit{e.g.}, stalling time and location, bitrate, etc.), which are challenging to obtain in advance in real-world scenarios, making them unsuitable for real-time live streaming. Moreover, these models do not validate their performance on live streaming specific distortions, raising doubts about their effectiveness in assessing the QoE of live video streaming.
\end{itemize}

To address these issues, we construct a first-of-its-kind live video streaming QoE assessment database, called the TaoLive QoE database, which investigates how authentic live streaming distortions affect users' QoE. Specifically, we collect $42$ source live streaming videos from the Taobao Live APP and artificially introduce various potential live streaming distortions, including compression artifacts, stalling distortions, accelerated frame rates, and frame skipping, to obtain $1,155$ distorted videos. Stalling events are induced by manipulating the presentation time stamp (\emph{PTS}) of the videos. All videos in our database are subjectively tested by a sufficient number of users to obtain thorough retrospective QoE scores. 

Subsequently, we benchmark existing QoE assessment models on the TaoLive QoE database. The results reveal that these models exhibit subpar performance on the TaoLive QoE database, whether they are reference-based QoE methods or rely on QoS features that are often inaccessible in real live streaming scenarios. In earlier studies, related QoS features usually needed to be calculated and extracted first. They used special algorithms for this. Then, these features were fed into the prediction model. To address this limitation, this study proposes a new method called Tao-QoE. Tao-QoE is an innovative QoE assessment approach. It is based on a deep neural network (DNN). This method can directly predict QoE scores from video input. Our method uses PTS of the video. It can effectively and indirectly capture and combine traditional QoS statistical features~\cite{aroussi2012empirical}. This way, it avoids the need for separate algorithms to extract explicit QoS features. The proposed method includes a multi-scale spatial feature extraction module to quantify intra-frame distortions and a optical flow-guided motion feature extraction module to capture inter-frame distortions. What's more, a video pre-processing module is employed to identify stalling events, facilitating the QoE model in comprehensively accounting for quality switching dynamics and the occurrence of stalling events. We validate the proposed model on TaoLive QoE database, as well as other VoD-based QoE databases and general-purpose VQA databases, demonstrating that the proposed model exhibits strong QoE and general video quality assessment capabilities.


The main contributions are summarized as follows:
\begin{itemize}[leftmargin=*]
\item \textbf{We establish the TaoLive database, a large-scale live streaming QoE database.} We collect $42$ source live videos from real-world live platforms and subsequently introduce live streaming specific distortions, including compression artifacts, stalling events, and frame skipping, to obtain $1,155$ distorted videos. A total of $20$ participants were invited to take part in a subjective QoE experiment to derive the ground-truth QoE scores.
\item \textbf{We propose Tao-QoE, a DNN-based QoE model for predicting video QoE in live streaming scenarios.} TAO-QoE reconstructs the video sequence by using PTS. This process implicitly encodes QoS information. Then, TAO-QoE uses multi-scale spatial feature extraction. It also uses optical flow-guided motion feature extraction. Based on these steps, it outputs a QoE score. TAO-QoE achieves the best performance on the TaoLive database. It also performs best on several public QoE databases. 

\end{itemize}

\vspace{-3mm} 
\section{Related Work}
\subsection{QoE \& VQA Databases}


Over the past decades, lots of subjective QoE database have been developed to address QoE assessment problem, serving as test benchmarks and providing training data for the development of objective QoE assessment methods. Table ~\ref{tab:QoE database} illustrates common used QoE databases, including WaterlooSQoE series~\cite{waterloo1,waterloo2,waterloo3,waterloo4} and LIVE-NFLX series~\cite{live1,live2}. These databases encompass a wide range of multimedia stimuli, such as images and videos, spanning different resolutions, compression levels, and content types. They also incorporate intentionally impaired content to simulate various degradation scenarios like compression artifacts, stalling event, and quality adaptation. The comparison between the TaoLive QoE database and other QoE databases includes the WaterlooSQoE database~\cite{waterloo1,waterloo2,waterloo3,waterloo4} and LIVE-NFLX~\cite{live2} is shown in Table \ref{tab:QoE database}.

The LIVE‑Qualcomm database~\cite{LIVE-Qualcomm} provides mobile video distortions with corresponding subjective scores. CVD2014~\cite{CVD2014} offers a benchmark for evaluating no‑reference VQA algorithms under realistic distortions. KoNViD‑1k database~\cite{KoNViD-1k} contains 1,200 natural video sequences with diverse authentic impairments. VDPVE database~\cite{VDPVE} focuses on perceptual video enhancement and includes both reference and processed videos. LIVE‑VQC database~\cite{LIVE-VQC} presents a large‑scale study of perceptual quality in authentically distorted videos. MSU database~\cite{MSU}introduces a comprehensive database designed for benchmarking learning‑based video quality metrics, especially in compression scenarios. YouTubeUGC database~\cite{YouTubeUGC} collects user‑generated content to support video compression research. LIVE‑WC database~\cite{LIVE-WC} explores the prediction of quality for compressed videos that already contain pre‑existing distortions. LIVE-APV database~\cite{LIVE_APV} is made for live video quality research. It includes six common types of video problems: compression, aliasing, judder, flicker, frame drop, and interlacing. The comparison between the VQA database is shown in Table \ref{tab:VQA database}.

\begin{table*}
    \centering
    \caption{Comparison of QoE databases.}
    \label{tab:QoE database}
    \begin{small}
    \begin{tabular}{cccccccc}
        \toprule
        Database  & Source Videos & \begin{tabular}[c]{@{}c@{}}Content\\ type\end{tabular} & Frame rate & \begin{tabular}[c]{@{}c@{}}Stalling\\ manifestation\end{tabular}   & \begin{tabular}[c]{@{}c@{}}Frames\\ Skipping\end{tabular}   &\begin{tabular}[c]{@{}c@{}}Fast\\ playback\end{tabular}    & Total number \\
        \hline
        LIVE-NFLX-I & 14 & VoD    & invariable &frame duplication& \ding{55}  &\ding{55} & 112\\
        LIVE-NFLX-II & 15 & VoD   & invariable &frame duplication& \ding{55}  &\ding{55} & 420\\
        WaterlooSQoE-I & 20 & VoD  & invariable &frame duplication& \ding{55}  &\ding{55}  & 200\\
        WaterlooSQoE-II & 12 & VoD  & invariable &frame duplication& \ding{55}  &\ding{55}  & 588\\
        WaterlooSQoE-III & 20 & VoD  & invariable &frame duplication& \ding{55}  &\ding{55}  & 450\\
        WaterlooSQoE-IV & 5 & VoD   & invariable &frame duplication& \ding{55}  &\ding{55} & 1,350\\
        \hline
        \textbf{Ours}  & 42                & Live video   &variable     &reset PTS &\ding{51} &\ding{51} & 1,155 \\
        \bottomrule
    \end{tabular}
    \end{small}
    
    \vspace{-5pt}
\end{table*}

\begin{table*}
    \centering
    \caption{Comparison of VQA databases.}
    \label{tab:VQA database}
    \begin{small}
    \begin{tabular}{ccccccc}
        \toprule
        Database   & Resolution(fps)&Distortion Type  & Total number\\ 
        \hline
        LIVE-Qualcomm  & 1080P(30) & in-capture distortions(artifacts, color,exposure,focus,sharpness, etc.)& 208\\
        
        CVD2014  & VGA\&720P(10$\sim$31) & in-capture distortions(sharpness, graininess, darkness, color balance,etc.)& 234\\
        
        KoNViD-1k  & various(20,25,30) & natural video sequences with diverse authentic impairments& 1200\\

        VDPVE  & 720P(various) & video enhancement processing(color,brightness,contrast,deblurring,etc.)& 1211\\

        LIVE-VQC  & various(30) & no preset labels, real-world mixed distortions& 585\\
        
        MSU  & various(24$\sim$60) & compression artifacts from modern encoding standards(AVC,HEVC,etc.)& 1022\\

        YouTubeUGC  & various(30) & authentic UGC distortions (Noise, Banding, etc.)& 1500\\

        LIVE-WC  & various(various) & compression artifacts (blocking,blur,ringing,etc.)& 275\\

        LIVE-APV  & 1080P\&4K(30) &
        H.264 compression(compression,aliasing,judder,flicker,frame drop,etc.)& 315\\
        
        \bottomrule
    \end{tabular}
    \end{small}
    
    \vspace{-15pt}
\end{table*}

\vspace{-5mm} 

\subsection{QoE \& VQA Models}
Early QoE studies  attempt to fit a mathematical formula to map the transmission-related metrics (\textit{i.e.,} QoS features) into video QoE~\cite{ref-3,mok2011measuring,xue2014assessing,yin2015control}. However, the video QoE is influenced by multiple factors, including spatial quality, temporal quality, play smoothness, video quality switching, video stuttering, etc. These factors interact and collectively impact users' QoE. Therefore, relying solely on statistical features is insufficient to capture users' subjective QoE. To improve the capability of QoE models in assessing the impact of visual quality on overall QoE, an increasing number of studies have integrated video quality assessment (VQA) within the QoE assessment framework. 
Both Spiteri~\cite{spiteri2020bola} and Bentaleb~\cite{bentaleb2016sdndash} regard the average bitrate of the video experienced by the user and the duration of the stalling events as the influencing factors of QoE. Duanmu \textit{et al.}~\cite{waterloo1} propose a QoE method, called SQI, which combines the Full-Reference Video Quality assessment(FR VQA) algorithm with video stalling quantification features to predict QoE scores. They later improve the SQI algorithm and develop the KSQI algorithm, which takes video presentation quality (represented by VMAF~\cite{vmaf}), stalling event, and quality adaptation (switching between profiles) into consideration~\cite{duanmu2019knowledge}. Bampis \textit{et al.}~\cite{ALTAS} develop the video assessment of temporal artifacts and stalls (Video ATLAS) metric, which unifies modeling of video visual quality, stall-related features, and memory-related features of video. 

With the rapid advancement of deep learning technology, more and more researchers apply DNN such as convolutional neural network (CNN) and recurrent neural network (RNN) to the prediction of video QoE. For example, GCNN-QoE~\cite{GCNN-QoE} performs 1D-CNN to fuse the statistical features and then utilizes GRU for the QoE regression. DA-QoE~\cite{DA-QoE} leverages 2D-CNN as well as 3D-CNN to extract segment-level features and then a multi-task prediction framework is proposed to simultaneously regress continuous-time and retrospective QoE scores. DeSVQ~\cite{DeSVQ} feeds the high-level spatio-temporal features extracted by CNN and the temporal features  by LSTM~\cite{yu2019review} to LSTM in turn, and finally returns the QoE score. Chen \textit{et al.}~\cite{TRR-QoE} develop a temporal reasoning guided QoE method named TRR-QoE, which extracts frame-level semantic features by CNN and propose a multi-scale temporal relational reasoning module to fuse the spatial and temporal information.  . Li \textit{et al.}~\cite{luchunyi} propose a real-time blind QoE metric as ASPECT that employs a non-uniform sampling method to reduce the complexity of semantic features extracted by the CNN. The extracted semantic features as well as the QoS features are then regressed to the QoE score using Support Vector Regression (SVR)~\cite{awad2015support}.

To address the challenges of video quality assessment, various deep learning-based models have been proposed. VMAF~\cite{vmaf} is a widely used full-reference video quality metric. It combines several basic quality measures. It also uses machine learning. Its goal is to mimic how people subjectively score video quality. This metric shows strong correlation in modern video coding evaluations. It also performs well in adaptive streaming scenarios. LTVQM~\cite{LTVQM} introduces a two-level framework that combines spatiotemporal features with a regressor to predict perceptual quality scores. VSFA~\cite{VSFA} proposes a method for quality assessment of “in-the-wild” videos by jointly learning semantic and distortion features in an end-to-end manner. SimpleVQA~\cite{SimpleVQA} presents a lightweight yet effective model for user-generated content (UGC) videos, leveraging multi-scale feature aggregation and attention mechanisms. FastVQA~\cite{FastVQA} further improves efficiency with a fragment sampling strategy, enabling real-time quality evaluation without sacrificing accuracy.

\begin{figure}[htb]
\centering
\includegraphics[width=3.5in]{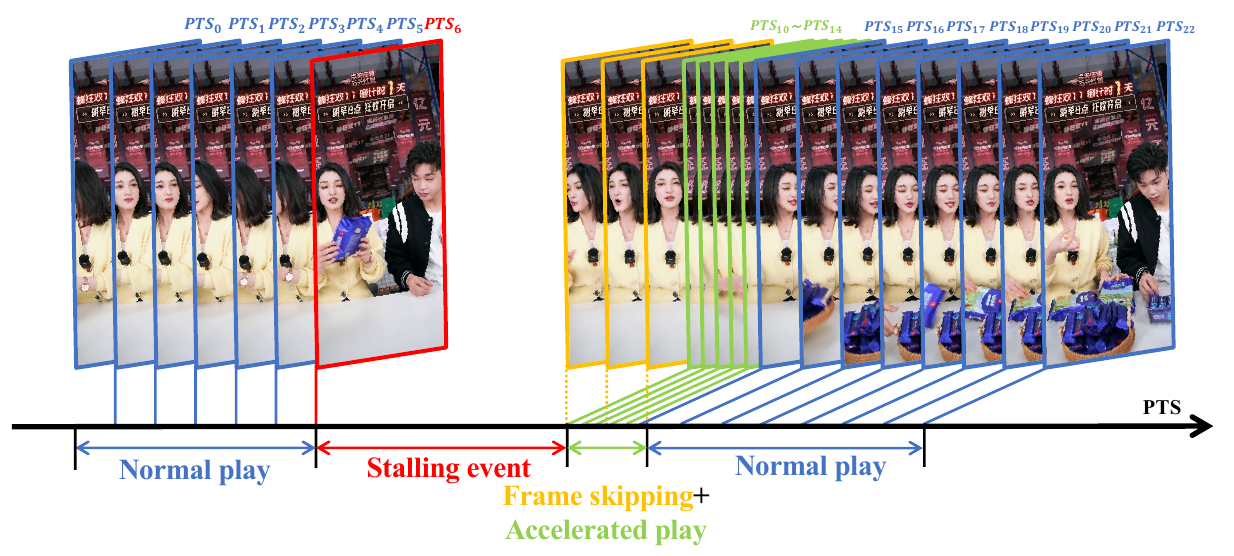} 
\caption{Stalling event, accelerated playing and frame skipping in TaoLive QoE Database. The blue video frames represent the frames played according to the source video frame rate, while the red video frames depict the displayed frames during stalling events. Additionally, green video frames indicate accelerated playback (accelerated frame rate), and yellow video frames signify skipped frames due to prolonged stalling duration.}
\label{video play}
\vspace{-15pt}
\end{figure}


\section{TaoLive QoE database}
Current QoE databases have several limitations when applied to live streaming scenarios: i) limited content diversity in source videos, particularly lacking human-related activities, while live streaming scenarios typically feature a wide range of such activities; ii) stalling events are predominantly represented as repeated frames, typically caused by network issues leading to uneven PTS distribution; iii) live streaming scenarios generally involve brief stalling periods, however, current QoE databases do not encompass stalling events lasting less than one second, a common occurrence in real-time broadcasts; iv) after such stalling events, live broadcasts often switch to accelerated playing with either an increased frame rate or frame skipping. These variations in frame rates are not addressed in publicly QoE databases that usually maintain a fixed frame rate.

To tackle these challenges, we present the TaoLive QoE database, which encompasses a larger corpus of representative source videos tailored for live streaming scenarios and incorporates real-world live streaming setups, including compression, accelerated frame rate playback, frame skipping, and more. 
Fig. \ref{video play} illustrates the presence of stalling events, accelerated frame rate playback, and frame skipping within the TaoLive QoE database. 



\vspace{-5mm} 
\subsection{Database Construction}
\label{sec:Database Constructio}
\subsubsection{Source Videos}
We carefully selected $42$ authentic live videos from the Taobao Live APP, covering various resolutions and frame rates. Specifically, the sources videos contains two resolutions (1080p and 720p) and three frame rates (20fps, 25fps, and 30fps). Specifically, for each combination of frame rate (20fps, 25fps, 30fps) and resolution (1080p, 720p), the number of corresponding source videos is consistently seven. Each video is 10 seconds long. To ensure optimal video performance, we excluded any videos with stalling events. 

\subsubsection{Live Streaming Distortions}
The distortions introduced in the TaoLive database include compression, stalling events, accelerated playing following a stalling event, and frame skipping. Due to the characteristics of live streaming, once a stalling event occurs, playback resumes with certain frames played at an accelerated rate. Frame skipping occurs when the stalling event duration surpasses a specific threshold. The speed and duration of accelerated play are primarily influenced by the the length of the stalling event. To simulate video transmission under different bandwidth conditions, we compress these source videos using FFmpeg with constant rate factors (CRF) of $15$, $22$, $27$, $32$, and $37$. 

Specifically, the $7$ source videos for each frame rate(20fps, 25fps, 30fps) and resolutions(1080p, 720p) are compressed based on the aforementioned $5$ CRFs. Subsequently, we manually introduce stalling events to these compressed videos. To ensure that no secondary compression occurs during the addition of stalling events, we utilize FFmpeg to modify PTS of the video in accordance with the designated stalling mode. This mode encompasses various combinations of stalling event duration and frequency. The duration of a stalling event is categorized into four levels: short (s) ($0.5$s or $1$s), medium (m) ($1.5$s, $2$s or $2.5$s), long (l) ($3$s, $3.5$s, $4$s or $4.5$s), and extra long (el) ($5$s, $5.5$s or $6$s). The maximum limit for stalling event occurrences is set to $3$. As depicted in Table \ref{21 stalling modes}, there are a total of $21$ combinations. The acceleration rate (AR) applied to expedite video playback following the termination of a stalling event is configured as $1.1$, $1.25$, $1.5$, $1.75$, and $2.25$. This parameter represents the ratio of the accelerated frame rate of video frames after a stalling event to the original frame rate.

\begin{table}[t]
\centering
\caption{21 stalling modes.. A\#, B\#, and C\# represent the identifiers for each stalling mode. The number of mode A, mode B, and mode C is 8, 8, and 5, respectively. A\#$\times 2$ indicates the presence of two occurrences of stalling mode A\#.For example, stalling mode C2 represents the artificial insertion of two short stalling events and one medium stalling event into the compress video.}
\label{21 stalling modes}
\begin{tabular}{|c|c|}
  \hline
  \# of Stalling  & Mode  \\
  \hline
  1 & \begin{tabular}[c]{@{}c@{}}1s (A1)$\times 2$, 1m (A2)$\times 2$,\\ 1m (A2)$\times 2$, 1l (A3)$\times 2$, 1el (A4)$\times 2$\end{tabular}   \\
  \hline
  2 & \begin{tabular}[c]{@{}c@{}}2s (B1), 1s + 1m (B2),  1s+1l (B3), 1s+1el (B4),\\ 2m (B5), 1m + 1l (B6), 1m + 1el (B7), 2l (B8)\end{tabular}  \\   
  \hline
  3 & \begin{tabular}[c]{@{}c@{}}3s (C1), 2s + 1m (C2), 2s + 1l (C3),\\ 1s + 2l (C4),1s + 1m + 1l (C5)\end{tabular}    \\   
  \hline
  
\end{tabular} 
\vspace{-15pt}
\end{table}

For stalling events, the specific generation procedure is described as follows. Let $F = \left\{f_1,f_2,...,f_N \right\}$ represent all the video frames of a compressed video $v$, and $P = \left\{p_1,p_2,...,p_N \right\}$ denote PTS corresponding to these frames $F$, where $n$ is the number of compressed video frames. Additionally, let $L = \left\{l_1,l_2,...,l_M \right\}$ represent the time points at which stalling events occur, and $T = \left\{t_1,t_2,...,t_M \right\}$ represent the durations of the stalling events, where $M$ is the total number of stalling events. First, the index of the stall video frame is calculated according to the occurrence time of the stalling event and the video frame rate. The indices of the stall video frames, denoted as $SF = \left\{sf_1,sf_2,...,sf_M \right\}$ are defined by

\begin{equation}
sf_j = l_j \times {\rm framerate} \quad j = 1,2...,M.
\end{equation}

Secondly, the PTS delay $D = \left\{d_1,d_2,...,d_M \right\}$ for all video frames after this frame is calculated by 

\begin{equation}
d_j = \frac{t_j}{ \rm timebase} \quad j = 1,2...,M,
\end{equation}
where ${\rm timebase}$ is the time base of $v$. The PTS delay of all video frames of the compressed video $AD = \left\{ad_1, ad_2, ..., ad_N\right\}$ is calculated as 

\begin{equation}
    ad_i = \begin{cases}
				0  \quad \quad i \le sf_1 \\
				\sum_{k} d_k  \quad sf_k < i \le sf_{k+1} \\
                \sum_{k}^{m} d_k  \quad i > sf_{M},
		\end{cases}
\end{equation}
where $i$ represents the frame index of the compressed video, adjustments to certain PTS values are necessary in order to ensure smooth playback following a stalling event. Specifically, the PTS interval for accelerated-playing video frames should be reduced based on the predetermined acceleration rate, while keeping the intervals for other frames unchanged. In the FFmpeg structure AVPacket, the PTS interval is represented as a constant (pkt.duration). The frame index, PTS, and pkt.duration of the video are calculated as $\rm PTS = index * pkt.duration$.The total number of accelerated video frames after a stalling event is directly proportional to the duration of the stall. This count reflects the cumulative number of accelerated frames needed to catch up with the live stream's progress delayed by the stall. The total number of accelerated-playing video frames, $QN = \left\{qn_1,qn_2,...,qn_M \right\}$, is given by

\begin{equation}
    qn_j = \frac{t_j \times AR \times {\rm framerate}}{AR - 1}\quad j = 1,2...,M.
\end{equation}

When accelerated playing, the PTS interval between the current and subsequent video frames is reduced according to the AR, while the PTS intervals of the remaining frames remain unchanged, equal to $pkt.duration$. 

To illustrate the acceleration playback mechanism, a video with a frame rate of 30 fps is used as an example. Assume that a stalling event with a duration of 0.5 seconds occurs during playback. If AR is set to 2, the system will accelerated-play the subsequent 30 video frames at a rate of 60 fps immediately after the stalling event concludes. During this accelerated-playing period, if another stalling event occurs, the accelerated-playing process is terminated immediately. Otherwise, after all 30 video frames have been accelerated-played, the video resumes normal playback at the original frame rate (30 fps).

Next, we recalculate the PTS values for the compressed video as $SP = \left\{sp_1, sp_2,...,sp_N\right\}$.  Finally we add the PTS delay of all video frames $AD = \left\{ad_1, ad_2, ..., ad_N\right\}$ and $SP = \left\{sp_1, sp_2,...,sp_N\right\}$ to get the PTS of the output compressed videos with stalling exvents. Algorithm 1~\ref{alg:The PTS calculation process of the output video} illustrates the process of converting the PTS sequence of a set of compressed video frames without stalling events into the PTS sequence after artificially adding stalling events.

All video sequences in the TaoLive QoE Database contain their original audio tracks. During subjective experiments, participants were presented with the videos including audio. For stalling events, the audio playback was synchronized with the video, i.e., audio was also paused during the stalling period and resumed afterward. It is worth noting that we did not introduce additional audio quality degradation or audio-video synchronization distortions in this study, as our focus is on the visual quality dimensions of live streaming.

\begin{algorithm}[tb]
\caption{The calculation process from the input compressed video's PTS to the output video's PTS}
\textbf{Input}: $P = \left\{p_N\right\}$; $SF = \left\{sf_M \right\}$; $D = \left\{d_M \right\}$; $AD = \left\{ad_N\right\}$; $QN = \left\{qn_M \right\}$; Total number of video frames: $N$; Number of stalling events: $M$; acceleration rate:$AR$;$pkt.duration$; \\
\textbf{Output}: $SP = \left\{sp_1, sp_2,...,sp_N\right\}$  \\
\begin{algorithmic}[1] 
\STATE Let $ QN\_end = [qne_1, qne_2,...,qne_M] = [0,...,0]$
\FOR{$i = 0; i < M-1; i++ $}
    \IF {$qn_i > sf_{i+1} - sf_i$}
    \STATE $qne_i = sf_{i+1} - 1$
    \ELSE
    \STATE $qne_i = qn_{i+1} + sf_i$
    \ENDIF
\ENDFOR
\IF {$i == M-1$}
        \IF{$qn_i > N - sf_i$}
        \STATE $qne_i = N - 1$
        \ELSE
        \STATE $qne_i = qn_{i} + sf_i$
        \ENDIF
    \ENDIF
\STATE The set of frame indexes that need to be accelerated $NAF = \left\{sf_1,...,sf_1 + qne_1,sf_m,...,sf_1 + qne_M \right\}$
\STATE $sp_1 = p_1$
\FOR{$i = 1; i < N; i++ $}
    \IF{$i \in NAF$}
        \STATE  $sp_i = sp_{i-1} + pkt.duration / AR $
    \ELSE
        \STATE $sp_i = sp_{i-1} + pkt.duration $
    \ENDIF
\ENDFOR

\end{algorithmic}
\label{alg:The PTS calculation process of the output video}

\end{algorithm}

\begin{table}[htb]
\caption{AR settings for videos of different resolutions. }
\label{AR}
\centering
\begin{tabular}{|c|c|c|c|c|c|c|}
  \hline
   & 1 & 1.1 & 1.25 & 1.5 & 1.75 & 2.25  \\
   \hline
  1080p(1st) & 100\% & - & - & - & - & - \\
  \hline
  1080p(2nd) & - & 30\% & 30\% & 15\% & 15\% & 15\% \\
  \hline
  720p(1st) & 25\% & 25\% & 20\% & 15\% & 10\% & 5\% \\   
  \hline
  
\end{tabular} 
\vspace{-5pt}
\end{table}

\subsubsection{Summary} 
For each of the 7 compressed videos (which have different CRFs, frame rates, and resolutions), three stalling events with different modes are introduced. To investigate the impact of accelerated video playback on QoE, for the initial batch of distorted 1080p videos with stalling events, the acceleration rate (AR) is set to 1. For the second batch of distorted 1080p videos and the first batch of 720p videos with stalling events, AR is randomly assigned based on the probabilities outlined in Table~\ref{AR}.  The occurrence of each stalling event was randomized.

\begin{figure}[t]
\centering
\includegraphics[width=3.in]{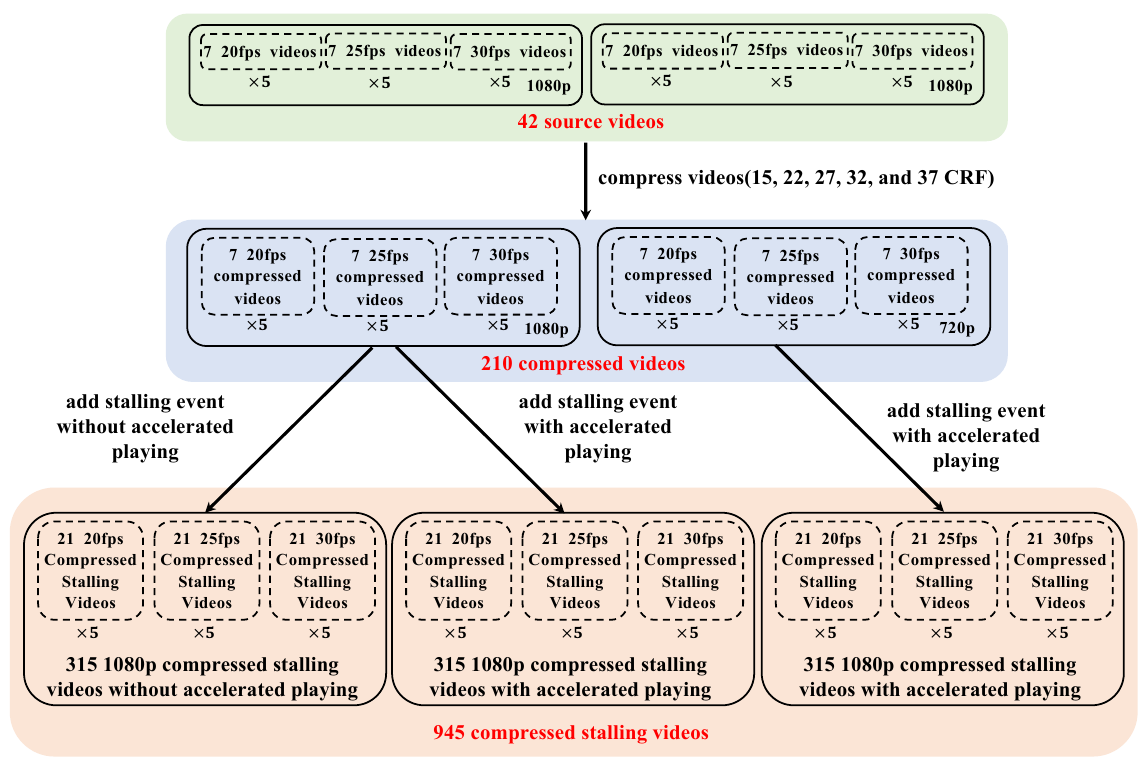}
\caption{The generation process for 945 compressed videos with stalling events}
\label{generation_process}
\vspace{-18pt}
\end{figure}

\begin{figure}[t]
\centering
\includegraphics[width=3.5in]{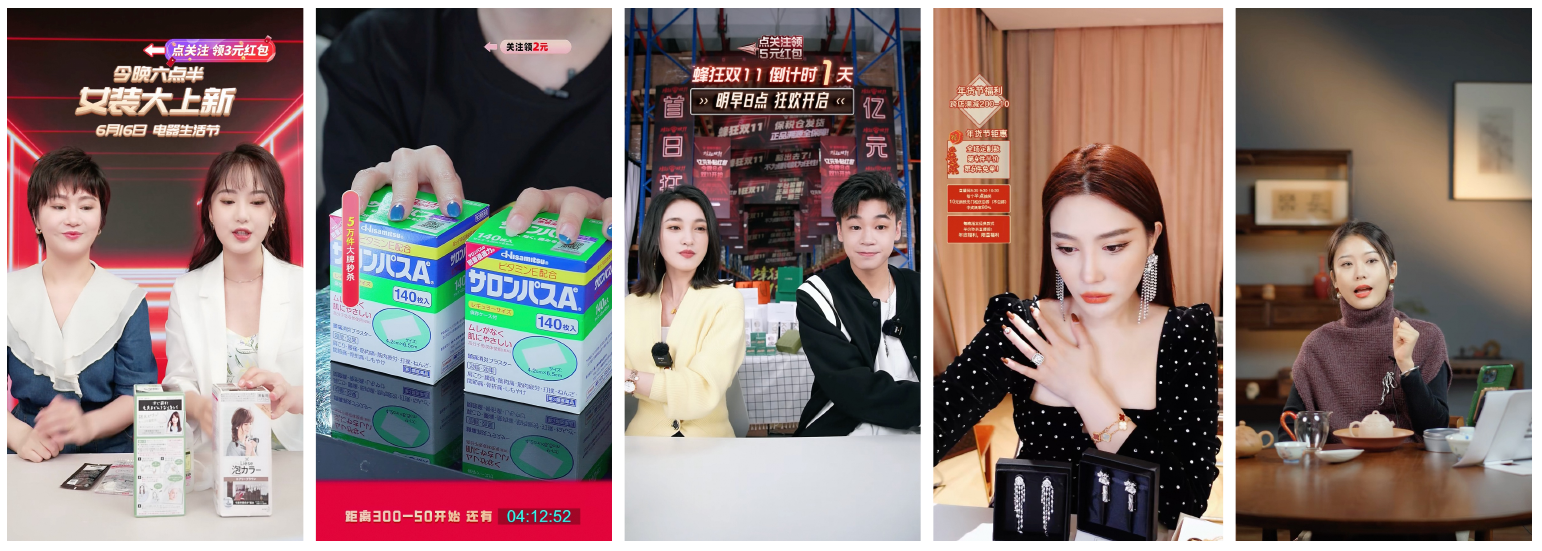}
\caption{Sample frames of the videos in the proposed TaoLive QoE Database.}
\label{TaoLive}
\vspace{-15pt}
\end{figure}

\begin{figure}[t]
    \centering
    \subfloat[Distribution for total video duration ]{\includegraphics[width=1.2in]{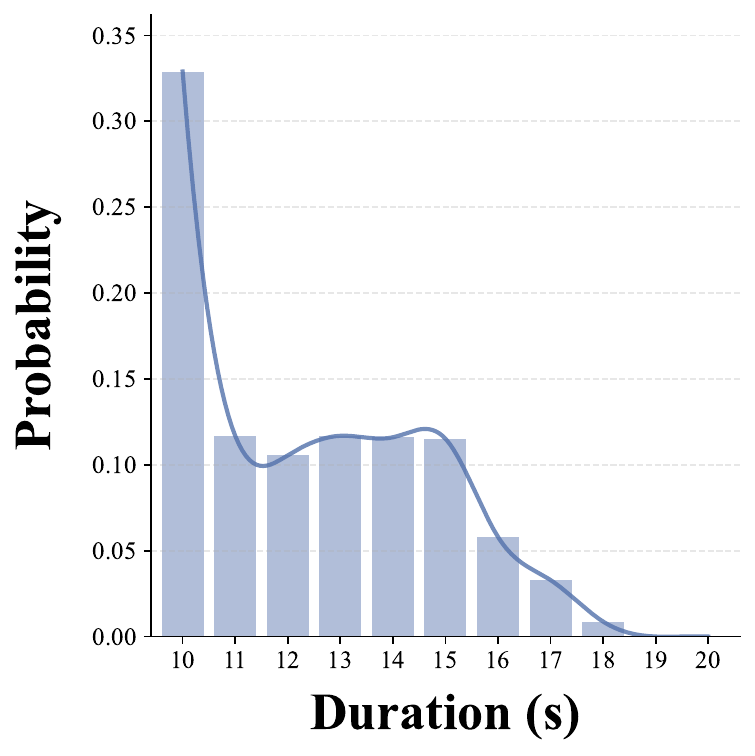}%
    \label{Duration_Distribution}}
    \hfil
    \subfloat[Distribution for total stalling duration/video duration duration ]{\includegraphics[width=2in]{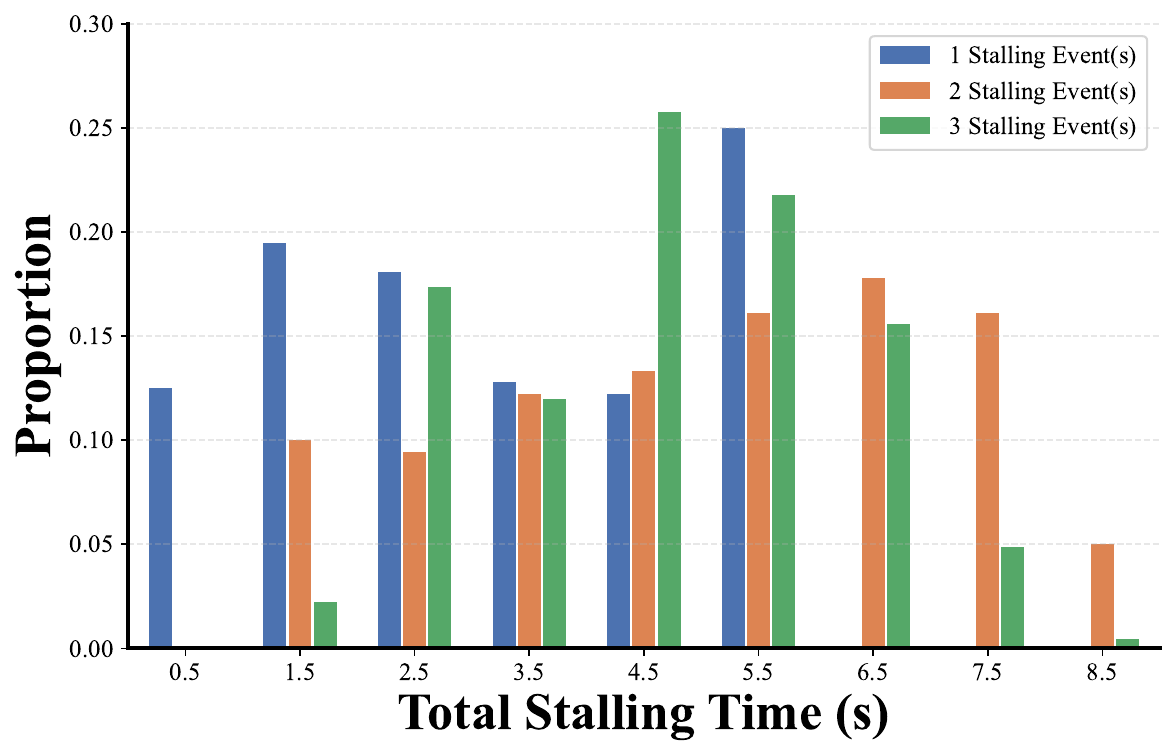}%
    \label{stalling_time_percentage}}
    \caption{Distribution for total video durations and the proportion of total stalling event duration relative to the total video duration}
\vspace{-15pt}
\end{figure}

A total of 945 videos with stalling events (21 1080p compressed videos $\times$ 3 stalling modes per source video $\times$ 5 CRFs $\times$ 2 batchs + 21 720p compressed videos $\times$ 3 stalling modes per source video $\times$ 5 CRFs $\times$ 1 batch) are generated. Fig.~\ref{generation_process} illustrates the process of generating 945 compressed videos with stalling events from 42 source videos. Additionally, $210$ videos without any stalling events are generated. Therefore, the complete database consists of $1155$ videos, with sample videos illustrated in Fig.~\ref{TaoLive}. In the complete database consisting of 1155 videos, the number (and percentage) of videos exhibiting 0, 1, 2, and 3 stalling events are 210 (18.18\%), 360 (31.17\%), 360 (31.17\%), and 225 (19.48\%), respectively. Fig.~\ref{Duration_Distribution} illustrates the distribution of total video durations across the entire database. Fig.~\ref{stalling_time_percentage} illustrates the distribution of the proportion of total stalling event duration relative to the total video duration.

\begin{figure}[t]
    \centering
    \subfloat[Distribution for confidence interval widths]{\includegraphics[width=1.8in]{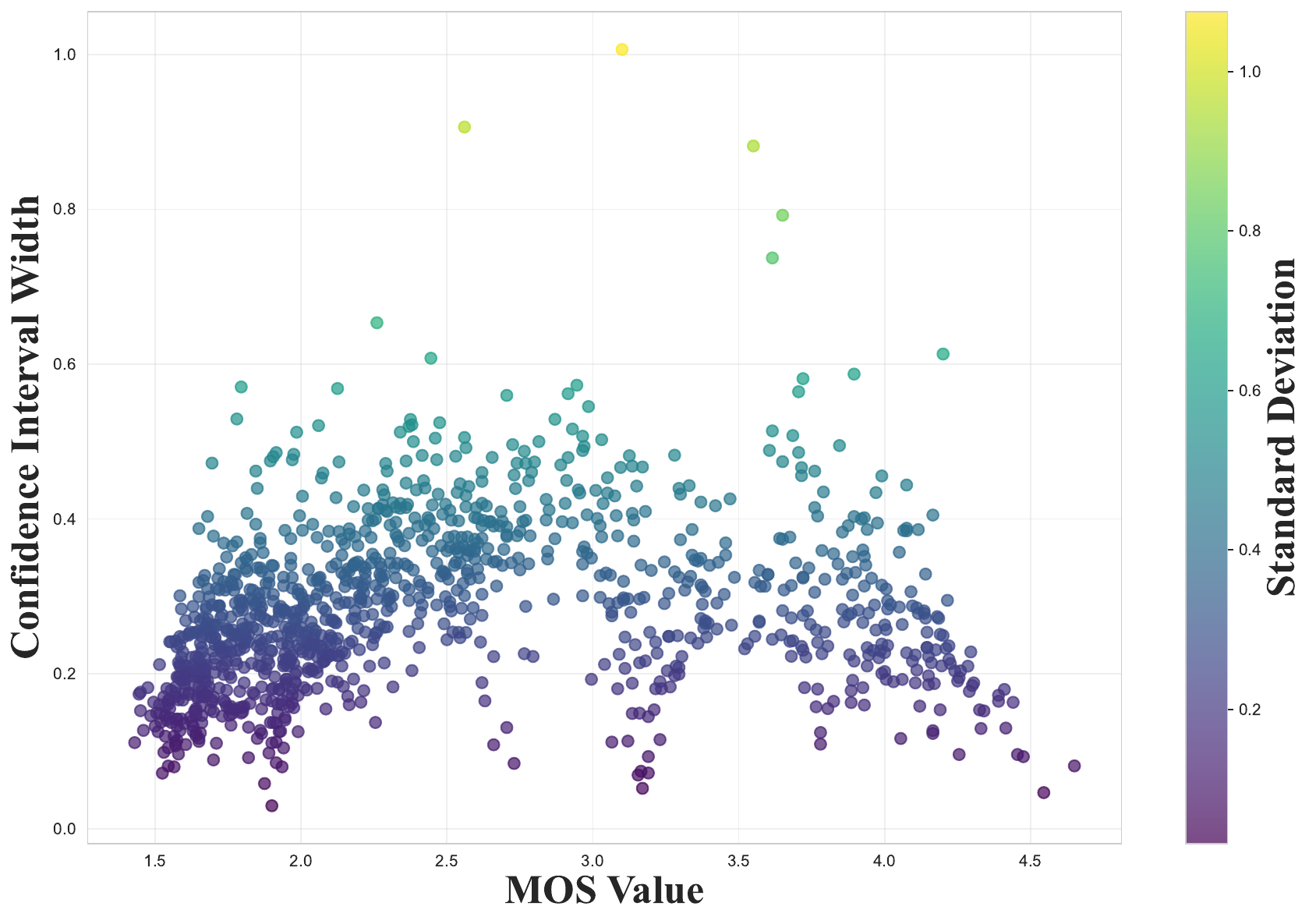}%
    \label{distribution_confidence_nterval_widths}}
    \hfil
    \subfloat[Percentages for confidence interval widths]{\includegraphics[width=1.3in]{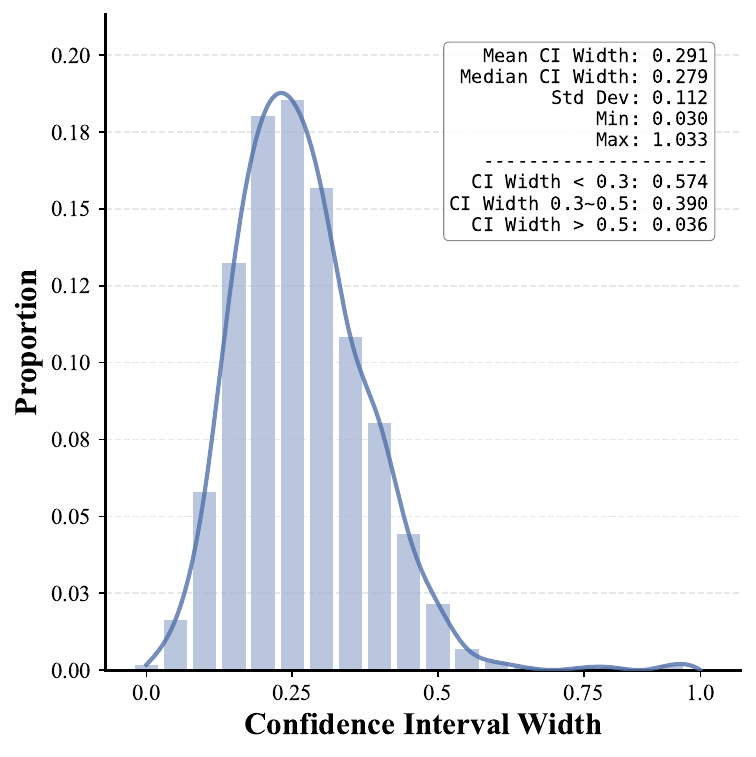}%
    \label{CI_width_distribution}}
    \caption{Distribution for confidence interval widths and their percentages}
\vspace{-15pt}
\end{figure}

\begin{figure}[t]
    \centering
    \subfloat[Distribution for PLCC of subject scores divided into two groups]{\includegraphics[width=1.5in]{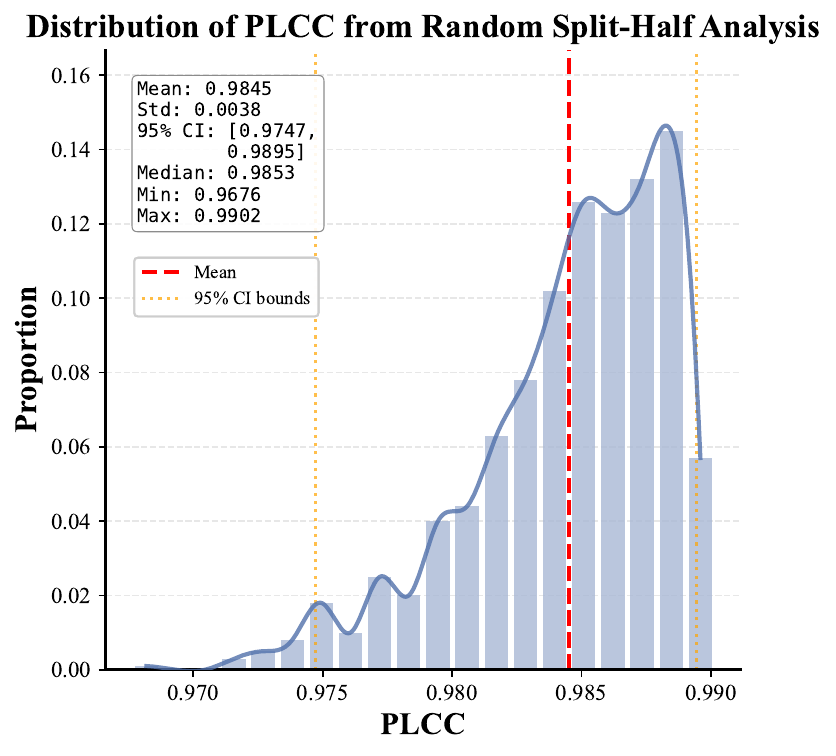}%
    \label{PLCC_distribution_1000_iteration}}
    \hfil
    \subfloat[Distribution for SRCC of subject scores divided into two groups]{\includegraphics[width=1.5in]{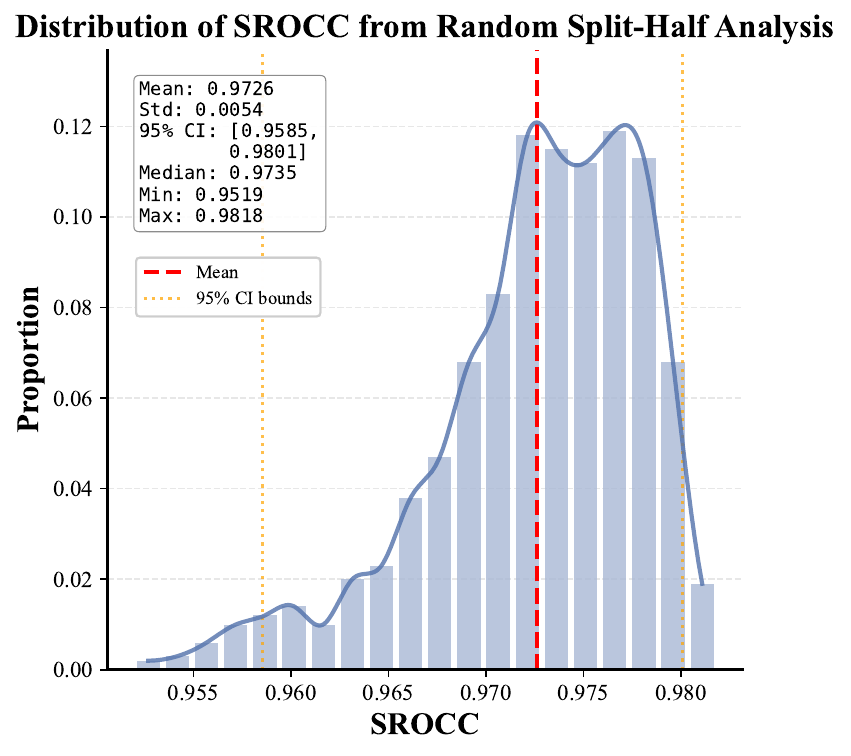}%
    \label{SROCC_distribution_1000_iteration.pdf}}
    \caption{Correlation distribution of subject scores divided into two groups for 1000 times}
\vspace{-15pt}
\end{figure}

\begin{figure*}[h]
    \centering
    \subfloat[MOS distributions]{\includegraphics[width=1.1in]{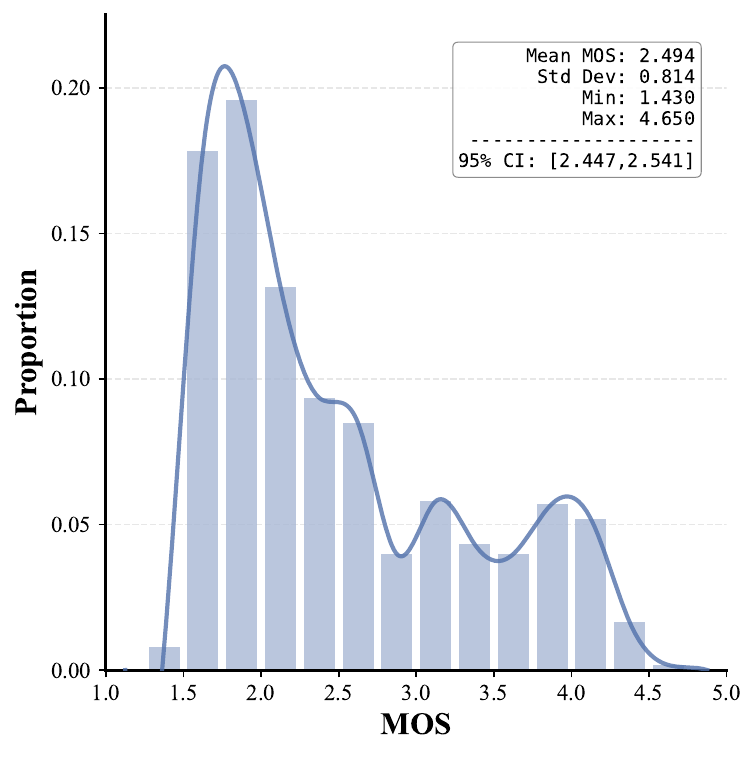}%
    \label{MOS distributions}}
    \hfil
    \subfloat[MOS distributions for different resolutions of no stalling video]{\includegraphics[width=1.4in]{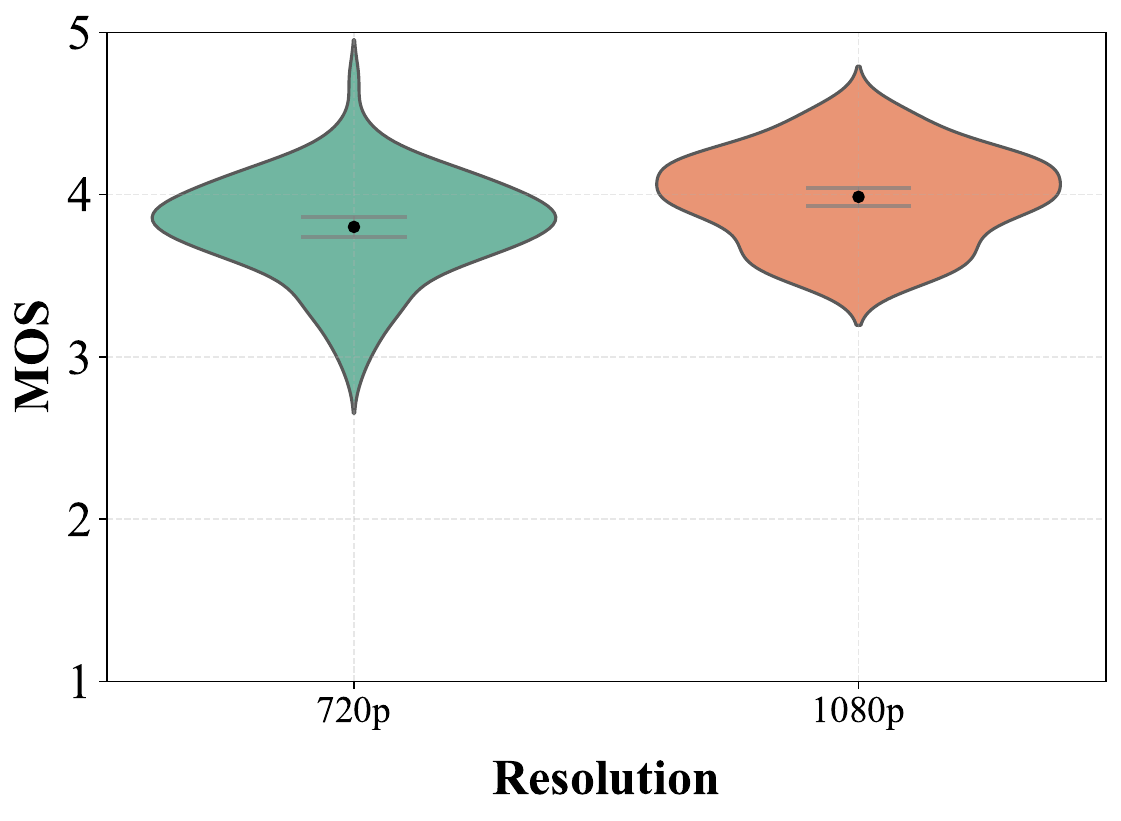}%
    \label{resolutions}}
    \hfil
    \subfloat[MOS distributions for different frame rates of no stalling video]{\includegraphics[width=1.4in]{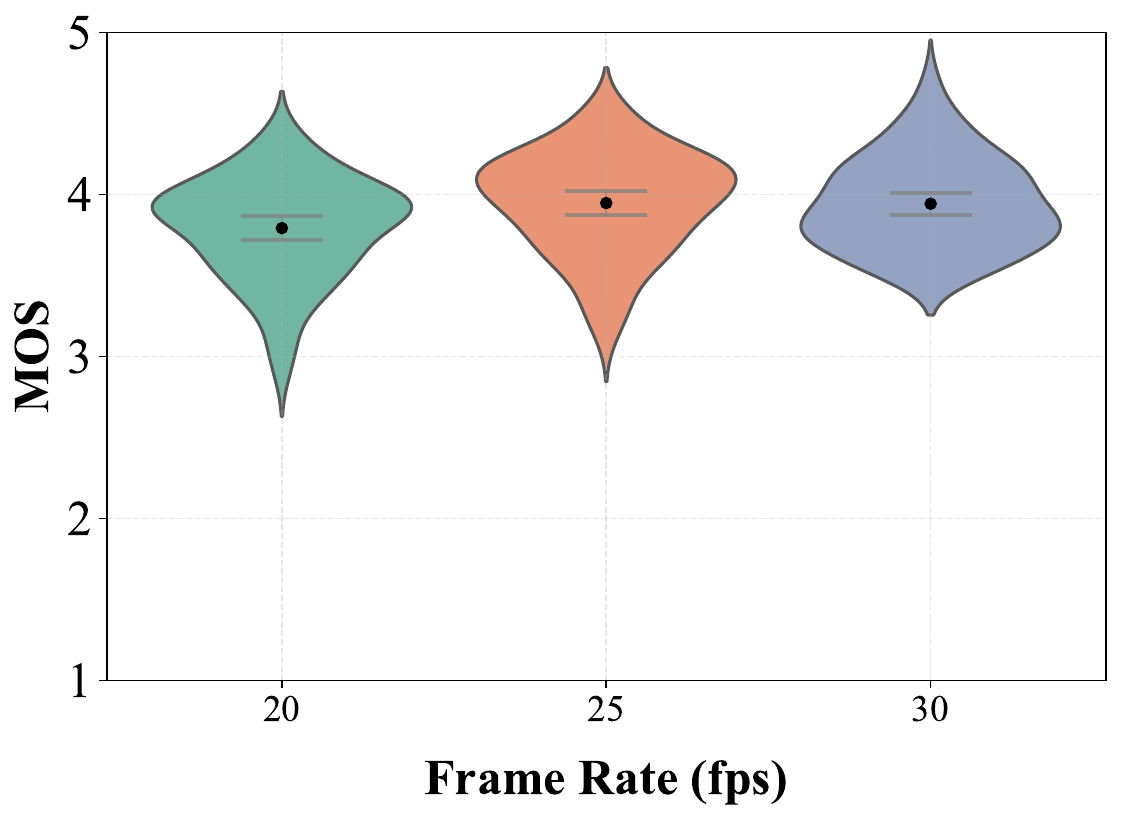}%
    \label{frame rates}}
    \hfil
    \subfloat[MOS distributions for different CRF values of no stalling video]{\includegraphics[width=1.4in]{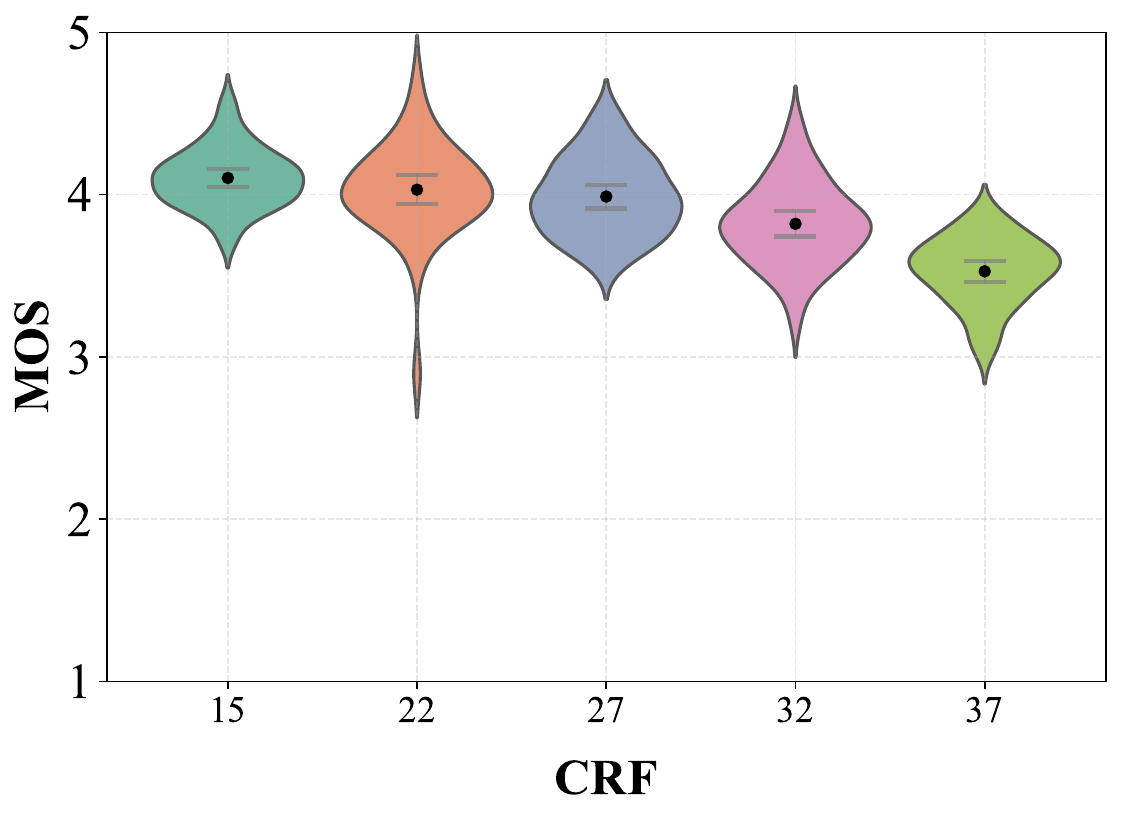}%
    \label{CRF}}
    \hfil
    \subfloat[MOS distributions for different stalling events numbers]{\includegraphics[width=1.4in]{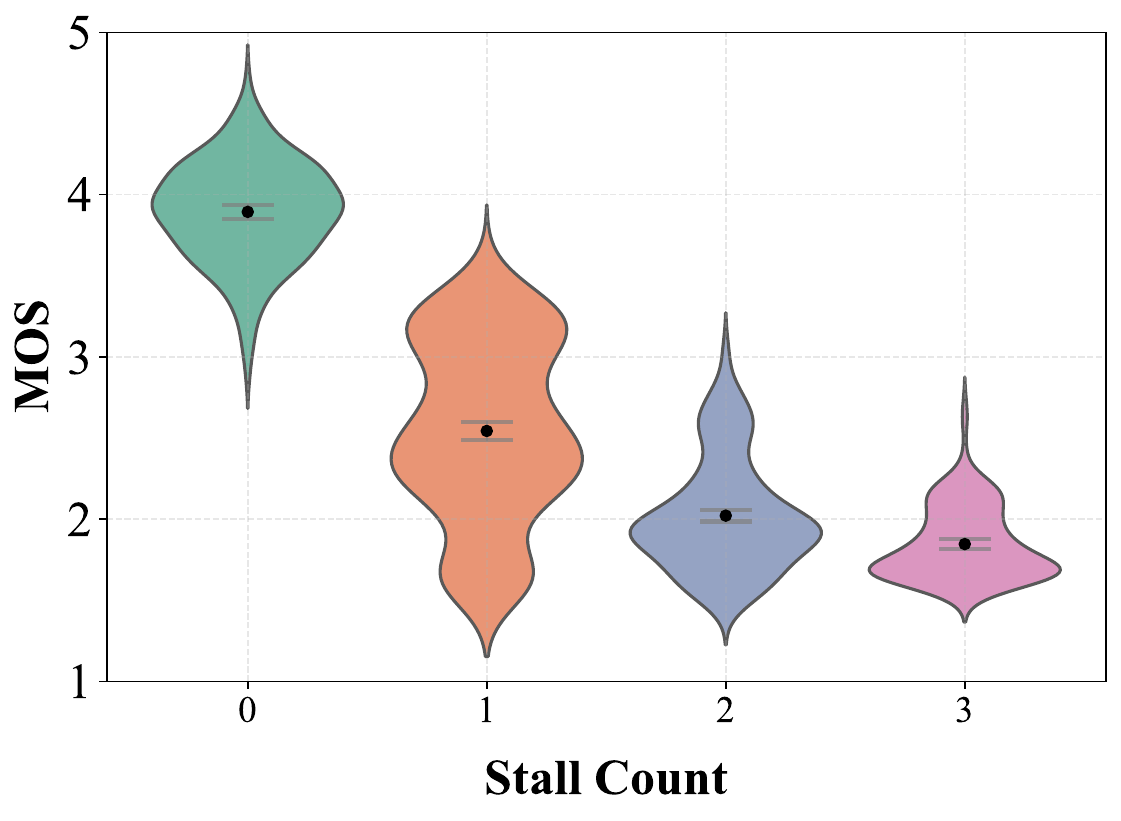}%
    \label{stalling events numbers}}
    \hfil
    \subfloat[MOS distributions for different ARs in different stalling modes]{\includegraphics[width=2.5in]{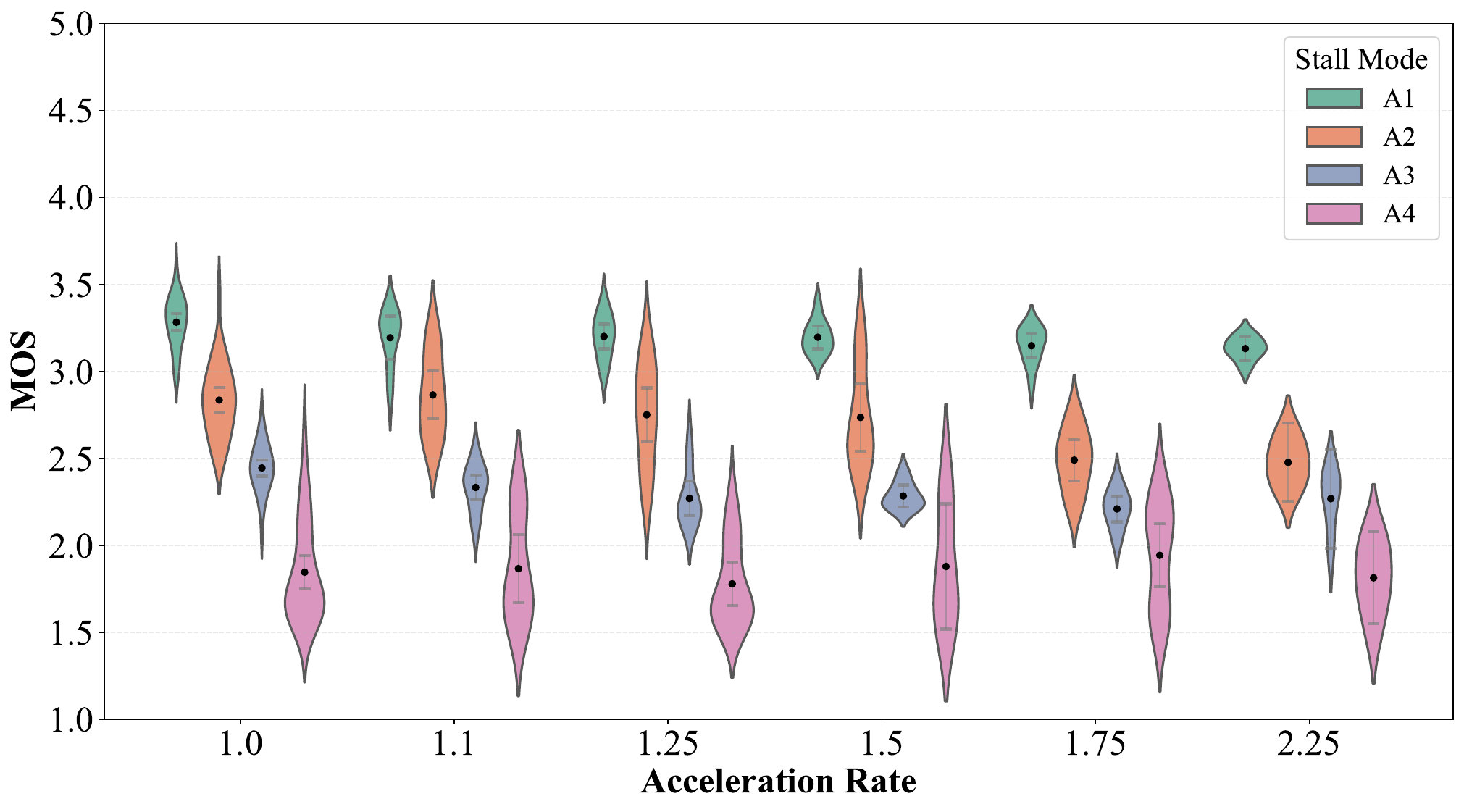}%
    \label{stalling modes}}
    \hfil
    \subfloat[MOS distributions for different stalling total durations]{\includegraphics[width=2.5in]{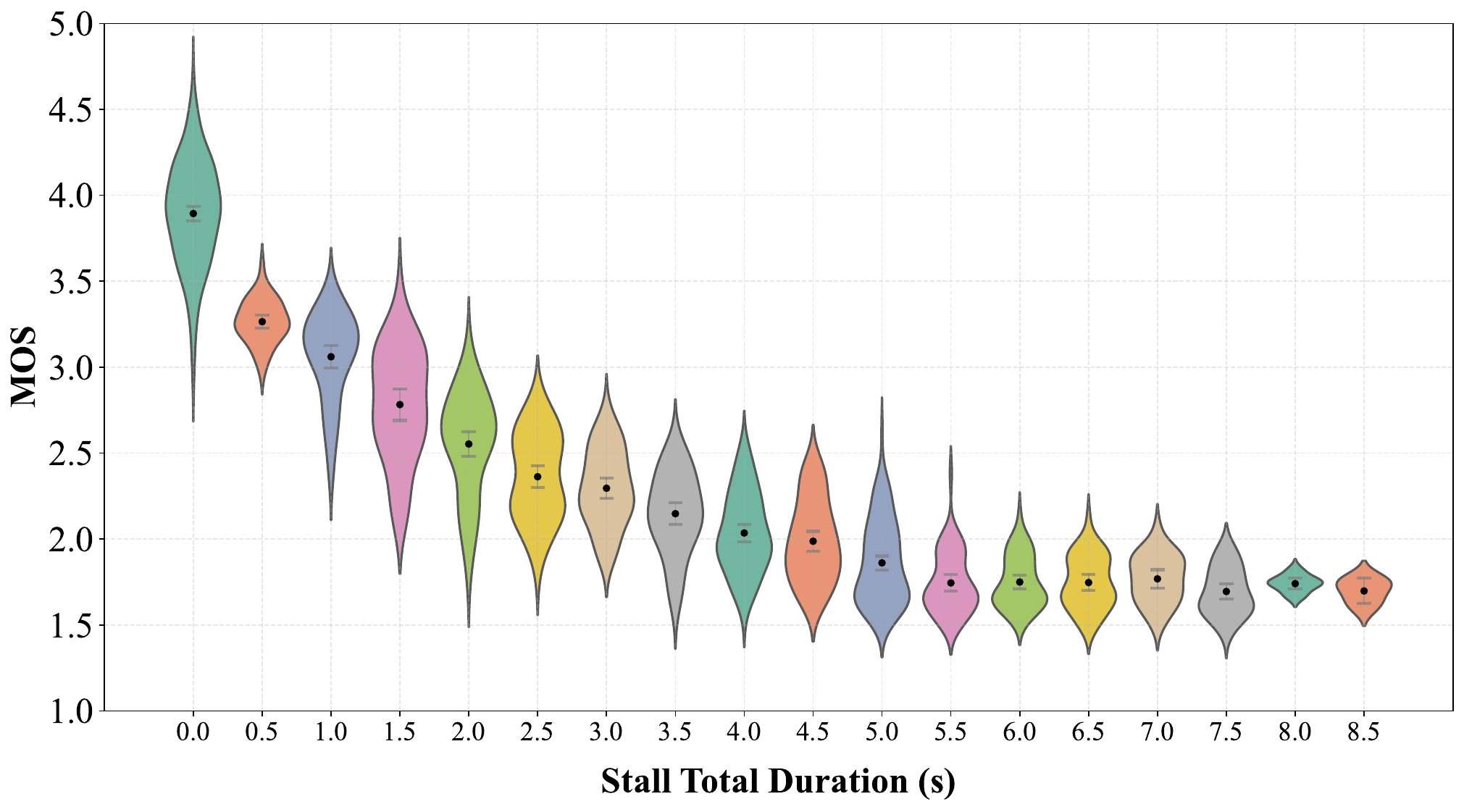}%
    \label{stalling total durations}}

    \caption{Illustration of the proposed TaoLive QoE database’s MOS distributions from different perspectives}
    \label{different perspectives}
\vspace{-15pt}
\end{figure*}

\vspace{-5mm} 

\subsection{Subjective Experiment Methodology} 
In the following, we present the detailed comprehensive methodology and configuration of the subjective test.

\begin{itemize}[leftmargin=*]
    \item \textbf{Subjective Testing Methodology:} ITU-R BT.500-11~\cite{ITI} outlines several subjective testing methods for evaluating visual quality, including Single-Stimulus (SS), Double-Stimulus Impairment Scale (DSIS), and Paired Comparison (PC). In this study, we employed the SS method, as it aligns with the short duration of the videos and the need for retrospective scoring.

    \item \textbf{QoE Rating:} The QoE scores range from $1$ to $5$, reflecting the overall viewing experience, with higher scores indicating better visual quality.
    \item \textbf{Participants:} According to ITU-T P.910~\cite{p910}, we employ subject post-screening with outlier ratio and Spearman Rank Correlation Coefficient (SROCC). Annotations that differ by more than two levels from the mode of all annotations are considered potential outliers and will be meticulously removed by the experts. A subject whose rated scores with SROCC $<$ 0.8 or outlier ratio $>$ 2\% will be marked as unreliable and will be rejected. Ultimately, 20 subjects are retained. The subjects are aged between 22 and 30 years, consisting of 11 males and 9 females. All subjects have normal or corrected-to-normal visual acuity, and are non-expert viewers with no professional background in image/video processing or quality assessment. In our subjective evaluation experiment, the scoring process lasted for 15 days. To effectively reduce visual and mental fatigue among the subjects, we adopted a strategy of dividing the task into batches and groups. Specifically, we randomly divided the full database used into 8 batches. Each batch contained about 140 videos. The videos in each batch were further split evenly into two smaller groups. In terms of scheduling, subjects completed the scoring for videos within the same batch on different days, with a gap of one day between sessions. For scoring videos from different groups within the same batch, there was a break of at least one hour.
    

    
    \item \textbf{Test Device:} We developed a Python-based graphical user interface (GUI) capable of rendering videos according to the specified PTS and frame rate, while automatically collecting subjective quality scores. To prevent geometric distortion from scaling operations, videos were played at their original resolution with gray borders filling the surrounding area. The GUI was run on a computer equipped with a 2.4 GHz Intel Core i5 processor and 16 GB of RAM. The viewing setup featured a 24-inch ViewSonic VA2452SM display(1920 $\times$ 1080p, 99\% sRGB, 60Hz).
    
\end{itemize}

\section{Data Processing and Analysis}
According to ITU-T P.910~\cite{p910}, We use the Mean Opinion Score (MOS) as the QoE label for each video, and its calculation formula is as follows:
\begin{equation}
MOS_{j} = \frac{1}{N} \sum_{i=1}^{N} r_{ij},  
\end{equation}

where $MOS_{j}$ represents the MOS for the j-th video,
N is the number of the valid annotation scores rated by subjects, and $r_{ij}$ is the rated score of the i-th subject on the j-th
video.

 


 

We calculated 95\% confidence intervals for the rating data of all videos. The specific distribution of confidence interval widths and their percentages are shown in the Fig. \ref{distribution_confidence_nterval_widths} and \ref{CI_width_distribution}. Looking at the distribution of interval widths, 57.4\% of videos have a width in the range [0, 0.3). Also, 39.0\% of videos have a width in the range [0.3, 0.5]. Only 3.6\% of videos have a width exceeding 0.5. This distribution indicates that the ratings given by different subjects are highly consistent. Therefore, the reliability of the rating data is good.

We conducted an inter-rater reliability correlation analysis. Specifically, we randomly divided the 20 subjects into two groups. For each group, we calculated the Mean Opinion Score (MOS) for all 1155 videos. Then, we computed the Spearman's Rank-Order Correlation Coefficient (SROCC) and the Pearson's Linear Correlation Coefficient (PLCC) between the MOS values of the two groups. To improve statistical reliability, we repeated this random grouping and correlation calculation process 1000 times. The distributions of the calculated PLCC and SROCC after 1000 groupings are shown in the Fig. \ref{PLCC_distribution_1000_iteration} and \ref{SROCC_distribution_1000_iteration.pdf}.

Figure \ref{different perspectives} illustrates the MOS distributions of the proposed TaoLive QoE database from multiple perspectives. Fig. \ref{MOS distributions} illustrates the distribution probability of MOS in the TaoLive QoE Database. The MOS distribution in this database is predominantly skewed toward lower scores, which aligns with the expected quality performance of videos containing stalling events. This characteristic makes the database well-suited for investigating the impact of stalling events on QoE. 

As shown in Fig.\ref{resolutions}, based on a 95\% confidence interval analysis, the user rating for 1080P resolution (MOS: 3.931–4.041) is significantly higher than for 720P resolution (MOS: 3.741–3.861). The fact that their confidence intervals do not overlap confirms a statistically significant difference (p $<$ 0.05). The average increase of 0.185 points indicates that users can perceive (or expect) a visual improvement from higher resolution.  

As shown in Fig.\ref{frame rates}, increasing the frame rate from 20 fps to 25 fps leads to a statistically significant improvement in experience. However, when the frame rate is further increased to 30 fps, its confidence interval clearly overlaps with that of 25 fps. This pattern shows a perceptual saturation effect. Improving from 20 fps to 25 fps effectively enhances perceived quality. In comparison, further increasing to 30 fps provides only limited marginal benefit. 

As shown in Fig. \ref{CRF}, stronger compression significantly reduces perceived quality. At lower CRF levels (15 to 22), the quality loss is not yet significant. However, when CRF exceeds 27, quality degradation accelerates and statistical differences become clear. In practical applications, it is recommended to keep CRF between 22 and 27. This balances quality and compression efficiency. A CRF higher than 32 will lead to a noticeable drop in experience.

As depicted in Fig. \ref{stalling events numbers}, the results reveal a negative correlation between stalling distortion and QoE scores, indicating that an increase in stalling events leads to a decline in QoE. Specifically, when the number of stalling events doubles, most videos receive QoE scores below 2. Moreover, with a threefold increase in stalling events, the proportion of videos scoring below 2 further increases.

As shown in Fig.~\ref{stalling modes}, we analyzed MOS scores under different acceleration ratios (AR) and stalling modes (A1-A4). As AR increases from the baseline value of 1.0 up to 2.25, user experience scores show a gradual overall decline. However, the decrease is relatively mild. Most adjacent confidence intervals overlap. This indicates that within the AR range of 1.0 to 2.25, increasing the video playback speed (i.e., accelerated playing) does not cause a statistically significant negative impact on the overall experience (p $\geq$ 0.05). At the same time, there is a clear hierarchy in experience among different stalling modes. The experience under mode A1 (which has the highest scores) is significantly better than under modes A2, A3, and A4. The confidence intervals between these modes basically do not overlap. This shows that the stalling mode is a key factor affecting the experience. In comparison, the adjustment effect of accelerated playing is relatively limited. This result suggests that, in scenarios where stalls occur, optimizing the stalling mode (for example, by reducing the duration of each stall) may be more important for improving QoE than adjusting the playback speed.


As shown in Fig.~\ref{stalling total durations}, as total stalling duration increases, MOS scores decline significantly, with the initial stalling event causing the most severe drop in user experience. Each additional second of stalling event leads to a statistically significant deterioration, though the rate of decline slows after about 4 seconds, suggesting a lower tolerance limit is approached. These findings confirm that total stalling duration strongly predicts streaming QoE, highlighting the critical importance of minimizing initial stalling event to preserve viewing quality.

\begin{figure*}[htb]
\centering
\includegraphics[width=1.0\textwidth]{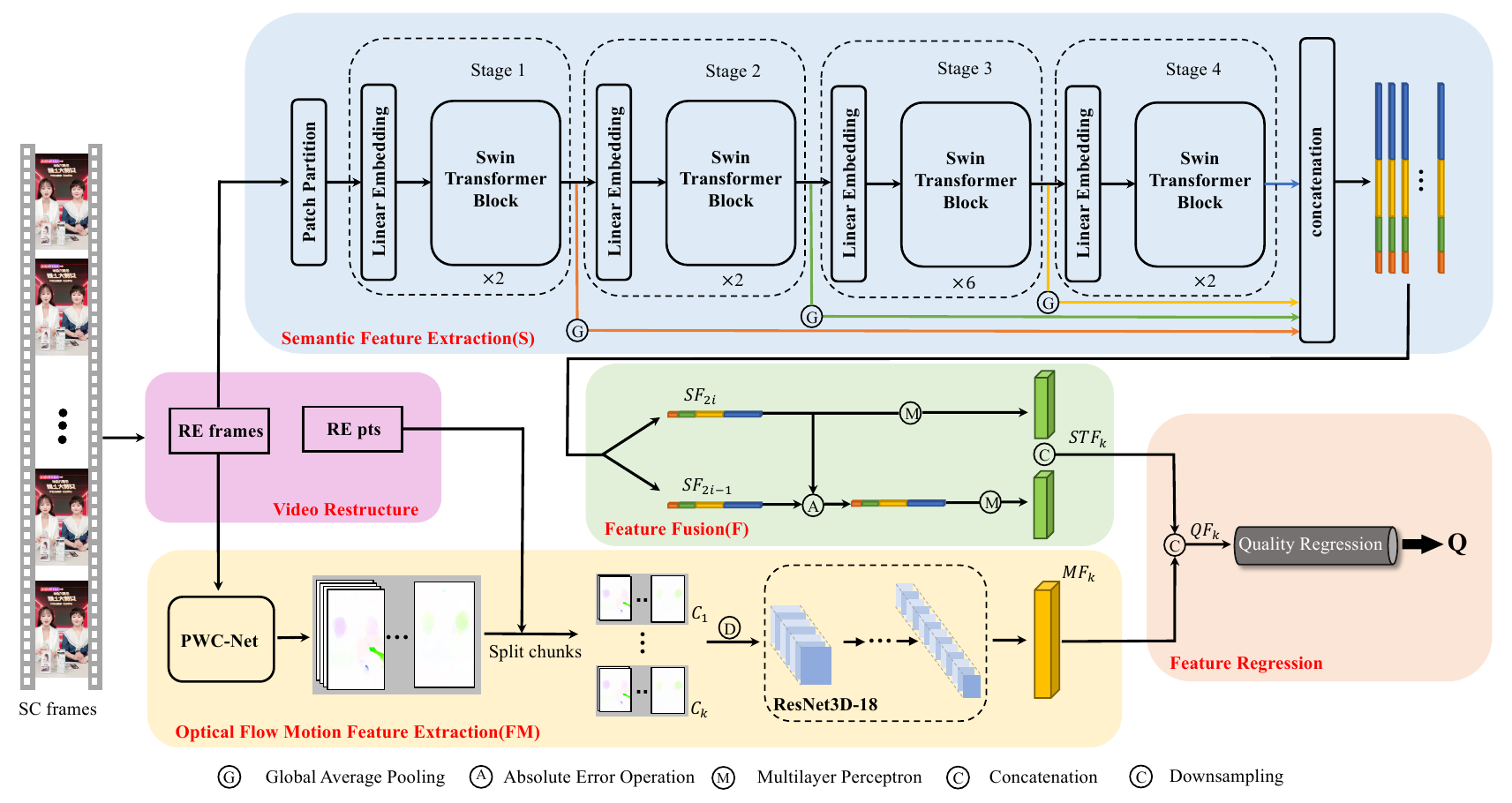} 
\caption{The overall structure of the proposed network. 1)semantic feature extraction module to extract semantic features from individual input frames; 2)optiacl flow motion feature extraction module to extract optiacl flow motion features information between frames; 3)multi-scale feature fusion module to process the extracted quality features; 4)feature regression module to predict retrospective QoE score.}
\label{fig2}
\vspace{-15pt}
\end{figure*}

\section{Proposed Method} 
The network architecture comprises five modules: a video reconstruction module, a semantic feature extraction module, an optiacl flow motion feature extraction module, a multi-scale feature fusion module, and a feature regression module, as illustrated in Fig.~\ref{fig2}. When evaluating a distorted video, the video reconstruction module first analyzes the input frames to detect stalling events caused by discontinuous PTS or accelerated playing. If such events are identified, it supplements the corresponding stalling frames based on the video's frame rate and the duration of the stall, thus generating the restructured video frame sequence along with its corresponding presentation timestamps (PTS). The semantic feature extraction module extracts the quality-aware features of all restructured video frames. Next, the multi-scale feature fusion module fuse the extracted quality-aware features. The motion feature extraction module then captures optical flow-based motion features from the restructured video frames. Finally, the feature regression module predicts retrospective QoE scores by integrating information from both quality and motion aspects.

\subsection{Video Reconstruction Module}
Mathematically, given a video consists of $N$ input frames $F = \left\{f_1,f_2,...,f_N \right\}$. We use FFmpeg to compute the theoretical PTS interval $\hat{P}$ between frames, as well as the PTS values of all input frames $P = \left\{p_1,p_2,...,p_N \right\}$. If the actual PTS interval between frame $i$ and frame $i+1$ exceeds the theoretical value $\hat{P}$,we identify a stalling event between these frames, with frame $i$ marked as the stalling frame. The model duplicates the stalling frame. The number of repeated reads, denoted as $rn$, is calculated as:

\begin{equation}
rn = \lfloor \frac{p_{i+1}-p_i}{\hat{P}} \rfloor
\end{equation}

where  $p_i$ indicates the PTS of frame $i$ and $\hat{P}$ indicates theoretical PTS interval between frames. 
We summarize the procedure for restructuring the video in Algorithm~\ref{alg:video restructure}. We obtained the Reference Frames (RE frames) $V = \left\{v_1, v_2,...,v_M\right\}$. The RE frames are then downsampled at 16fps to ensure the model can accept videos with varying frame rates. The downsampled video frames $V = \left\{v_1,v_2,...,v_{2L} \right\}$ serve as input to semantic Feature extraction module.

\begin{algorithm}[htb]
\caption{video restructure}
\label{alg:video restructure}
\textbf{Input}: Frames of input video $F = \left\{f_1,f_2,...,f_N \right\}$; Total number of input video frames $N$ \\
\textbf{Output}: RE frames $V = \left\{v_1, v_2,...,v_M\right\}$; PTS of RE frames $REP = \left\{rep_1, rep_2,...,rep_M\right\}$  \\
\begin{algorithmic}[1] 
\STATE Read  $\hat{P} = pkt.duration$ by FFmpeg
\STATE Read  PTS of input video frames $P = \left\{p_1,p_2,...,p_N \right\}$ by FFmpeg
\FOR{$i = 1; i < N; i++ $}
    \IF {$\hat{P} >= p_{i+1} - p_i$}
    \STATE $v_i = f_i, rep_i = p_i$
    \ELSE
    \STATE $v_i = f_i, rep_i = p_i, rn = \lfloor \frac{p_{i+1}-p_i}{\hat{P}} \rfloor$
    \FOR{$j = 0; j < rn; j++ $}
        \STATE $v_{i+j} = f_i, rep_{i+j} = p_i + \hat{P} * j$
    \ENDFOR
    \ENDIF
\ENDFOR
\STATE $v_M = f_N, rep_m = p_N$

\end{algorithmic}

\end{algorithm}

\subsection{Semantic Feature Extraction Module} 

We employ the pre-trained Swin Transformer~\cite{swin} as the backbone network architecture. The primary goal of the semantic feature extraction network is to capture multi-scale semantic features for each frame. Notably, different semantic content can influence human tolerance to various distortions in distinct ways~\cite{li2019quality}. Furthermore, incorporating semantic information can aid in detecting and measuring perceptual distortions, making it a reasonable addition to the quality assessment. Additionally, video quality is hierarchical in nature, with perception occurring from low-level features to high-level ones. To account for this hierarchy, we concatenate the multi-scale features extracted across the four stages of the Swin Transformer and use them as frame-level semantic features.

Specifically, we feed these downsampled video frames $V = \left\{v_1,v_2,...,v_{2L} \right\}$ into the semantic feature extraction network. $SF = \left\{SF_1,SF_2,...,SF_{2L} \right\}$ is the output multi-scale semantic features.

\begin{equation}
\begin{split}
SF_i = \alpha_1 \oplus \alpha_2 \oplus \alpha_3 \oplus \alpha_4  \quad i = 1,2...,2L \\
\alpha_j = GAP(L_j(v_i))  \quad j = 1,2,3,4
\end{split}
\end{equation}

where  $SF_i$ indicates the extracted semantic features from $i$-th frame $v_i$. $GAP(\cdot)$ represents the global average pooling operation. $L_j(v_i)$ is the feature of the $j$-th stage output of Swin Transformer. $\alpha_j$ denotes the average pooled features from $L_j(v_i)$.

\subsection{Optiacl Flow Motion Feature Extraction Module} 

Live broadcasts are often affected by unstable shooting environments and restricted network conditions, resulting in motion distortions such as jitter, stalling events and accelerated playing. Therefore, relying solely on semantic features at the frame level is insufficient to accurately capture these distortions. 
Hence, it is imperative to incorporate motion features in QoE prediction models. To detect stalling events effectively, we 
extract inter-frame optical flow images using a pretrained PWC-Net~\cite{sun2018pwc}. Subsequently, we segment the extracted optical flow based on PTS of Reference Frames (RE frames), with each segment having a duration of 1 second. 

\begin{equation}
C_k = \Gamma(V)  
\end{equation}
where $\Gamma(\cdot)$ represents the operation of extracting optiacl flow, and $C_k$ represents the extraction and clipping operations of inter-frame optical flow images for RE frames. We employ PTS to perform the clipping operation on inter-frame optical flow images. In case of accelerated video playback, the number of optical flow images contained in $C_k$ may vary.

Subsequently, the inter-frame optical flow images are resampled at a rate of 16fps for each clip, followed by leveraging a pre-trained 3D-CNN backbone ResNet-18~\cite{resnet} to capture motion distortion at the clip level.

\begin{equation}
MF_k = \Phi(C_k)  
\end{equation}

The optiacl flow motion features extracted from clip $C_k$ are denoted as $MF_k$, and $\Phi(\cdot)$ represents the operation of extracting optiacl flow motion features.


\subsection{Multi-scale Feature Fusion Module} 
The evidence from~\cite{narwaria2012low} demonstrates that there exists an inverse relationship between video quality and adaptation quality, where lower adaptation quality contributes to a more enjoyable viewing experience for the audience. Consequently, the absolute error of semantic features between consecutive frames can serve as an indicator of adaptation quality.

\begin{equation}
SF_{2m}^{'} = |SF_{2m} - SF_{2m-1}|  \quad m = 1,2...,L,
\end{equation}
where $SF_{2m}^{'}$ represent the absolute error between adjacent semantic features. Then the multi-scale fusion can be derived as:

\begin{equation}
STF_{k} = W(\varphi(SF_{2m}) \oplus \varphi(SF_{2m}^{'}))   \quad m = 1,2...,L,
\end{equation}
where $\oplus(\cdot)$ stands for the concatenation operation, $\varphi(\cdot)$ represents the learnable Multilayer Perceptron. $W$ is a learnable linear mapping operation, and we finally obtain the spatio-temporal fused features $STF_k$.  Then we connect the spatio-temporal fusion feature and the optiacl flow motion feature to get the final QoE feature.

\begin{equation}
QF_{k} =  STF_k \oplus \varphi(MF_k)   
\end{equation}

\subsection{Feature Regression Module} 
After the aforementioned process of feature extraction and fusion, we employ a fully connected layer to perform regression on the QoE features in order to obtain QoE scores.

\begin{equation}
Q_{k} =  FC(QF_k)   
\end{equation}

where $FC(\cdot)$ is the fully-connected layers and $Q_i$ presents the QoE score of clip $C_k$. Finally, we average all clips of the input video to obtain a retrospective QoE score for that video.

\begin{equation}
Q =  \frac{r}{n} \sum_{1}^{\frac{n}{r}} Q_k 
\end{equation}

where $Q$ is the video QoE score and $\frac{n}{r}$ stands for the number of clips. We simply use the Mean Squared Error (MSE) as the loss function:

\begin{equation}
Loss =  \frac{1}{n} \sum_{i=1}^{n} (Q_g - Q_p)^2 
\end{equation}

where $n$ indicates the number of videos in a mini-batch, $Q_g$ and $Q_p$ are the MOS and predicted retrospective QoE score respectively.

\section{Experiment} 
In this section, we first provide a comprehensive description of the experimental setup. Subsequently, we evaluate the performance of our proposed Tao-QoE model and compare it with other prominent QoE models using our TaoLive QoE Database as well as publicly available QoE and VQA databases. Furthermore, ablation experiments are conducted to investigate the individual contributions of different sub-modules towards enhancing the overall model performance.

\subsection{Implementation Details}
The Tao-QoE model is implemented in PyTorch~\cite{pytorch}, with the Swin Transformer~\cite{swin} backbone utilizing pretrained weights from the ImageNet-1K ~\cite{imagenet} for semantic feature extraction. Additionally, the ResNet3D-18~\cite{resnet} pretrained on the Kinetics-400 ~\cite{imagenet} is adopted as the feature extraction network for the motion feature extraction module. The weights of both the multi-scale feature fusion module and the feature regression module are initialized randomly. The semantic feature extraction module processes input video frames at their original resolution ($1080 \times 1920$ or $960 \times 1280$). Meanwhile, the optiacl flow motion feature extraction module first extracts optical flow maps from the video frames at their original resolution, then resizes the maps to $224 \times 224$, and finally feeds them into a ResNet-18 3D-CNN. Our model was trained and tested on a server equipped with an Intel(R) Xeon(R) Platinum 8163 CPU @ 2.50GHz, 128GB RAM, and NVIDIA Tesla V100 SXM2.  The Adam optimizer~\cite{kingma2014adam} is utilized with an initial learning rate of $0.001$. In case the training loss fails to decrease within $5$ epochs, the learning rate is halved. The default number of epochs is set to $50$. During the process of optiacl flow motion feature extraction, all videos are down-sampled to a frame rate of 16fps for ensuring consistent feature dimensions. Following standard practice, we split the database into train and test sets at an $80$\%-$20$\% ratio. To assess the stability of the QoE model, we randomly perform $10$ content-based splits and record their average result as the final performance. Specifically, for the WaterlooSQoE-IV database, we performed content-based splitting five times due to the limited availability of only 5 source videos in the database. It is important to note that, to ensure fairness in the evaluation, video reconstruction was performed for all compared models. To ensure fairness in evaluation, we performed frame duplication for the video reconstruction module in all other compared models.

\subsection{Benchmark Databases \& Compared Models}


In the field of QoE, we selected TaoLive QoE  and five other available QoE databases, including LIVE-NFLEX-II~\cite{live2}, WaterlooSQoE-I~\cite{waterloo1}, WaterlooSQoE-II~\cite{waterloo2}, WaterlooSQoE-III~\cite{waterloo3} and WaterlooSQoE-IV~\cite{waterloo4} as the test benchmark.  We compare the proposed model with the following QoE models:

\begin{itemize}[leftmargin=*]
    \item Traditional models: P.1203~\cite{P1203}, SQI~\cite{duanmu2019knowledge}, Bentaleb~\cite{spiteri2020bola}, Spiteri~\cite{bentaleb2016sdndash}, VideoATLAS~\cite{ALTAS}, KSQI~\cite{duanmu2019knowledge}
    \item Deep learning-based models: GCNN-QoE~\cite{GCNN-QoE}, ASPECT~\cite{luchunyi}
\end{itemize}

Unfortunately, the code for the GCNN-QoE model is not publicly available, thus hindering our ability to assess its performance on TaoLive QoE .

Our model demonstrates strong predictive capabilities not only in the domain of QoE assessment, but also exhibits equally competent performance in the field of VQA. In the domain of VQA, we selected $9$ UGC VQA databases: LIVE-Qualcomn~\cite{LIVE-Qualcomm}, CVD2014~\cite{CVD2014}, KoNViD-1k~\cite{KoNViD-1k}, VDPVE~\cite{VDPVE}, LIVE-VQC~\cite{LIVE-VQC}, MSU~\cite{MSU}, YouTubeUGC~\cite{YouTubeUGC}, LIVE-WC~\cite{LIVE-WC},  LIVE-APV~\cite{LIVE_APV}.

We compare the proposed method with the following no-reference VQA models: LTVQM~\cite{LTVQM}, VSFA~\cite{VSFA}, SimpleVQA~\cite{SimpleVQA}, FastVQA~\cite{FastVQA}.

\subsection{Criteria}
Two types of evaluation criteria are employed to assess the performance of the models. The first, known as the Video Quality Experts Group (VQEG) criteria~\cite{VQEG-1, VQEG-2, VQEG-3}, evaluates the correlation between predicted scores and Mean Opinion Scores (MOS). The second, proposed by Krasula \textit{et al.}~\cite{criteria-2-1, criteria-2-2, criteria-2-3, criteria-2-4}, assesses the models' ability to determine whether two videos have similar quality scores or which one is superior. We refer to the former as the VQEG criteria and the latter as the classification criteria.

For the VQEG criteria, the model's predicted scores must first be mapped using the following function:

\begin{equation}
f(p) =  \xi_1(\frac{1}{2} - \frac{1}{1 + e^{\xi_2(p - \xi_3)}}) + \xi_4 p + \xi_5
\end{equation}

where $\left\{\xi | i = 1,2,3,4,5\right\}$ is the parameter to be fitted, $p$ and $f(p)$ represent the prediction score and mapping score respectively. The mapped scores are then used to calculate four correlation values with MOSs, namely Spearman Rank-Order Correlation Coefficient (SRCC), Pearson Linear Correlation Coefficient (PLCC), Root Mean Squared Error (RMSE), and Kendall Rank-order Correlation Coefficient (KRCC). These statistical indices serve different purposes in assessing model performance. Specifically, PLCC reflects the linearity of algorithm predictions, SRCC indicates their monotonicity or predictive correlation, while RMSE evaluates model consistency. An excellent model should achieve values close to 1 for SRCC, PLCC and KRCC.


For the classification criteria, we follow the procedures outlined in~\cite{VQEG-1} and apply the statistical methods from~\cite{Brill} to analyze subjective data, determining the significance of differences between each pair of stimuli. A $95$\% confidence level is employed. The entire  is divided into subsets based on significant differences and similarities. In significantly distinct subsets, stimulus pairs are further categorized by positive or negative MOS differences. Stronger model performance is indicated by a higher ability to distinguish between similar and dissimilar pairs, as well as superior and inferior stimuli. Therefore, we use the area under the ROC curve (AUC) as the evaluation metric for classification performance. Additionally, we compare the AUC values across different models to assess whether their performance differences are statistically significant~\cite{Hanley}.

We follow the VQEG standard to analyze the performance of the Tao-QoE model and other VQA models across different UGC s. However, since the VQA databases used in this study do not provide the standard deviation of annotation scores for each video, we are unable to apply the classification criteria. For QoE models, in line with standard practices~\cite{sunwei}, we employ both the VQEG criteria and classification criteria to evaluate the effectiveness of the Tao-QoE model and other QoE models on the QoE databases.

\begin{table*}[ht]
\centering
\caption{Comparison of QoE models. Best in {\color[HTML]{FE0000} \textbf{red}} and second in {\color[HTML]{3166FF} \textbf{blue}}}
\label{tab:QoE compare}
\resizebox{\textwidth}{!}{%
\begin{tabular}{cc|c|c|c|c|c|c|c|c|c|c|c|c}
\hline
\multicolumn{2}{c|}{{\color[HTML]{000000} Models}} & {\color[HTML]{000000} } & {\color[HTML]{000000} } & {\color[HTML]{000000} } & {\color[HTML]{000000} } & {\color[HTML]{000000} } & {\color[HTML]{000000} } & {\color[HTML]{000000} } & {\color[HTML]{000000} } & {\color[HTML]{000000} } & {\color[HTML]{000000} } & {\color[HTML]{000000} } & {\color[HTML]{000000} } \\ \cline{1-2}
\multicolumn{1}{c|}{{\color[HTML]{000000} Databases}} & {\color[HTML]{000000} Criteria} & \multirow{-2}{*}{{\color[HTML]{000000} Mok2011}} & \multirow{-2}{*}{{\color[HTML]{000000} FTW}} & \multirow{-2}{*}{{\color[HTML]{000000} Liu2012}} & \multirow{-2}{*}{{\color[HTML]{000000} Xue2014}} & \multirow{-2}{*}{{\color[HTML]{000000} P.1203}} & \multirow{-2}{*}{{\color[HTML]{000000} Bentaleb}} & \multirow{-2}{*}{{\color[HTML]{000000} Spiteri}} & \multirow{-2}{*}{{\color[HTML]{000000} SQI}} & \multirow{-2}{*}{{\color[HTML]{000000} KSQI}} & \multirow{-2}{*}{{\color[HTML]{000000} ASPECT}} & \multirow{-2}{*}{{\color[HTML]{000000} GCNN-QoE}} & \multirow{-2}{*}{{\color[HTML]{000000} Tao-QoE}} \\ \hline
\multicolumn{1}{c|}{{\color[HTML]{000000} }} & {\color[HTML]{000000} PLCC} & {\color[HTML]{000000} 0.614} & {\color[HTML]{000000} 0.470} & {\color[HTML]{000000} 0.794} & {\color[HTML]{000000} 0.761} & {\color[HTML]{000000} 0.654} & {\color[HTML]{000000} 0.903} & {\color[HTML]{000000} 0.790} & {\color[HTML]{000000} 0.904} & {\color[HTML]{000000} 0.863} & {\color[HTML]{000000} 0.667} & {\color[HTML]{3166FF} \textbf{0.935}} & {\color[HTML]{FE0000} \textbf{0.948}} \\
\multicolumn{1}{c|}{{\color[HTML]{000000} }} & {\color[HTML]{000000} SRCC} & {\color[HTML]{000000} 0.594} & {\color[HTML]{000000} 0.464} & {\color[HTML]{000000} 0.793} & {\color[HTML]{000000} 0.754} & {\color[HTML]{000000} 0.693} & {\color[HTML]{000000} 0.895} & {\color[HTML]{000000} 0.773} & {\color[HTML]{000000} 0.902} & {\color[HTML]{000000} 0.862} & {\color[HTML]{000000} 0.606} & {\color[HTML]{3166FF} \textbf{0.927}} & {\color[HTML]{FE0000} \textbf{0.946}} \\
\multicolumn{1}{c|}{{\color[HTML]{000000} }} & {\color[HTML]{000000} RMSE} & {\color[HTML]{000000} 0.610} & {\color[HTML]{000000} 0.501} & {\color[HTML]{000000} 0.456} & {\color[HTML]{000000} 0.468} & {\color[HTML]{000000} 0.565} & {\color[HTML]{3166FF} \textbf{0.322}} & {\color[HTML]{000000} 0.477} & {\color[HTML]{000000} 0.323} & {\color[HTML]{000000} 0.363} & {\color[HTML]{000000} 0.581} & {\color[HTML]{000000} /} & {\color[HTML]{FE0000} \textbf{0.250}} \\
\multicolumn{1}{c|}{\multirow{-4}{*}{{\color[HTML]{000000} LIVE-II}}} & {\color[HTML]{000000} KRCC} & {\color[HTML]{000000} 0.477} & {\color[HTML]{000000} 0.363} & {\color[HTML]{000000} 0.604} & {\color[HTML]{000000} 0.584} & {\color[HTML]{000000} 0.529} & {\color[HTML]{000000} 0.740} & {\color[HTML]{000000} 0.581} & {\color[HTML]{000000} 0.748} & {\color[HTML]{000000} 0.703} & {\color[HTML]{000000} 0.425} & {\color[HTML]{3166FF} \textbf{0.778}} & {\color[HTML]{FE0000} \textbf{0.800}} \\ \hline
\multicolumn{1}{c|}{{\color[HTML]{000000} }} & {\color[HTML]{000000} PLCC} & {\color[HTML]{000000} 0.478} & {\color[HTML]{000000} 0.488} & {\color[HTML]{000000} 0.596} & {\color[HTML]{000000} 0.781} & {\color[HTML]{000000} 0.561} & {\color[HTML]{000000} 0.920} & {\color[HTML]{000000} 0.834} & {\color[HTML]{000000} 0.799} & {\color[HTML]{000000} 0.909} & {\color[HTML]{000000} 0.790} & {\color[HTML]{FE0000} \textbf{0.945}} & {\color[HTML]{3166FF} \textbf{0.933}} \\
\multicolumn{1}{c|}{{\color[HTML]{000000} }} & {\color[HTML]{000000} SRCC} & {\color[HTML]{000000} 0.452} & {\color[HTML]{000000} 0.465} & {\color[HTML]{000000} 0.711} & {\color[HTML]{000000} 0.856} & {\color[HTML]{000000} 0.737} & {\color[HTML]{000000} 0.919} & {\color[HTML]{000000} 0.861} & {\color[HTML]{000000} 0.791} & {\color[HTML]{000000} 0.903} & {\color[HTML]{000000} 0.774} & {\color[HTML]{FE0000} \textbf{0.934}} & {\color[HTML]{3166FF} \textbf{0.929}} \\
\multicolumn{1}{c|}{{\color[HTML]{000000} }} & {\color[HTML]{000000} RMSE} & {\color[HTML]{000000} 17.552} & {\color[HTML]{000000} 17.152} & {\color[HTML]{000000} 14.285} & {\color[HTML]{000000} 9.962} & {\color[HTML]{000000} 12.865} & {\color[HTML]{3166FF} \textbf{7.250}} & {\color[HTML]{000000} 9.409} & {\color[HTML]{000000} 11.308} & {\color[HTML]{000000} 7.813} & {\color[HTML]{000000} 11.598} & {\color[HTML]{000000} /} & {\color[HTML]{FE0000} \textbf{7.163}} \\
\multicolumn{1}{c|}{\multirow{-4}{*}{{\color[HTML]{000000} WaterlooSQoE-I}}} & {\color[HTML]{000000} KRCC} & {\color[HTML]{000000} 0.363} & {\color[HTML]{000000} 0.369} & {\color[HTML]{000000} 0.528} & {\color[HTML]{000000} 0.679} & {\color[HTML]{000000} 0.558} & {\color[HTML]{000000} 0.758} & {\color[HTML]{000000} 0.683} & {\color[HTML]{000000} 0.615} & {\color[HTML]{000000} 0.738} & {\color[HTML]{000000} 0.594} & {\color[HTML]{FE0000} \textbf{0.806}} & {\color[HTML]{3166FF} \textbf{0.775}} \\ \hline
\multicolumn{1}{c|}{{\color[HTML]{000000} }} & {\color[HTML]{000000} PLCC} & {\color[HTML]{000000} 0.190} & {\color[HTML]{000000} 0.364} & {\color[HTML]{000000} 0.592} & {\color[HTML]{000000} 0.423} & {\color[HTML]{000000} 0.773} & {\color[HTML]{000000} 0.838} & {\color[HTML]{3166FF} \textbf{0.846}} & {\color[HTML]{000000} 0.685} & {\color[HTML]{000000} 0.691} & {\color[HTML]{000000} 0.803} & {\color[HTML]{000000} 0.826} & {\color[HTML]{FE0000} \textbf{0.874}} \\
\multicolumn{1}{c|}{{\color[HTML]{000000} }} & {\color[HTML]{000000} SRCC} & {\color[HTML]{000000} 0.173} & {\color[HTML]{000000} 0.305} & {\color[HTML]{000000} 0.595} & {\color[HTML]{000000} 0.433} & {\color[HTML]{000000} 0.801} & {\color[HTML]{000000} 0.818} & {\color[HTML]{3166FF} \textbf{0.820}} & {\color[HTML]{000000} 0.722} & {\color[HTML]{000000} 0.531} & {\color[HTML]{000000} 0.790} & {\color[HTML]{000000} 0.818} & {\color[HTML]{FE0000} \textbf{0.866}} \\
\multicolumn{1}{c|}{{\color[HTML]{000000} }} & {\color[HTML]{000000} RMSE} & {\color[HTML]{000000} 13.991} & {\color[HTML]{000000} 14.041} & {\color[HTML]{000000} 10.048} & {\color[HTML]{000000} 13.622} & {\color[HTML]{000000} 9.554} & {\color[HTML]{000000} 7.820} & {\color[HTML]{3166FF} \textbf{7.953}} & {\color[HTML]{000000} 10.766} & {\color[HTML]{000000} 10.307} & {\color[HTML]{000000} 9.665} & {\color[HTML]{000000} /} & {\color[HTML]{FE0000} \textbf{6.645}} \\
\multicolumn{1}{c|}{\multirow{-4}{*}{{\color[HTML]{000000} WaterlooSQoE-II}}} & {\color[HTML]{000000} KRCC} & {\color[HTML]{000000} 0.131} & {\color[HTML]{000000} 0.211} & {\color[HTML]{000000} 0.435} & {\color[HTML]{000000} 0.292} & {\color[HTML]{000000} 0.620} & {\color[HTML]{000000} 0.637} & {\color[HTML]{3166FF} \textbf{0.634}} & {\color[HTML]{000000} 0.531} & {\color[HTML]{000000} 0.383} & {\color[HTML]{000000} 0.605} & {\color[HTML]{000000} 0.624} & {\color[HTML]{FE0000} \textbf{0.680}} \\ \hline
\multicolumn{1}{c|}{{\color[HTML]{000000} }} & {\color[HTML]{000000} PLCC} & {\color[HTML]{000000} 0.302} & {\color[HTML]{000000} 0.423} & {\color[HTML]{000000} 0.606} & {\color[HTML]{000000} 0.481} & {\color[HTML]{000000} 0.782} & {\color[HTML]{000000} 0.855} & {\color[HTML]{000000} 0.820} & {\color[HTML]{000000} 0.723} & {\color[HTML]{000000} 0.682} & {\color[HTML]{000000} 0.798} & {\color[HTML]{3166FF} \textbf{0.890}} & {\color[HTML]{FE0000} \textbf{0.900}} \\
\multicolumn{1}{c|}{{\color[HTML]{000000} }} & {\color[HTML]{000000} SRCC} & {\color[HTML]{000000} 0.270} & {\color[HTML]{000000} 0.378} & {\color[HTML]{000000} 0.623} & {\color[HTML]{000000} 0.469} & {\color[HTML]{000000} 0.809} & {\color[HTML]{000000} 0.836} & {\color[HTML]{000000} 0.804} & {\color[HTML]{000000} 0.744} & {\color[HTML]{000000} 0.500} & {\color[HTML]{000000} 0.762} & {\color[HTML]{3166FF} \textbf{0.881}} & {\color[HTML]{FE0000} \textbf{0.890}} \\
\multicolumn{1}{c|}{{\color[HTML]{000000} }} & {\color[HTML]{000000} RMSE} & {\color[HTML]{000000} 13.444} & {\color[HTML]{000000} 13.459} & {\color[HTML]{000000} 9.989} & {\color[HTML]{000000} 12.828} & {\color[HTML]{000000} 9.219} & {\color[HTML]{3166FF} \textbf{7.364}} & {\color[HTML]{000000} 8.246} & {\color[HTML]{000000} 9.947} & {\color[HTML]{000000} 10.291} & {\color[HTML]{000000} 8.954} & {\color[HTML]{000000} /} & {\color[HTML]{FE0000} \textbf{6.494}} \\
\multicolumn{1}{c|}{\multirow{-4}{*}{{\color[HTML]{000000} WaterlooSQoE-III}}} & {\color[HTML]{000000} KRCC} & {\color[HTML]{000000} 0.204} & {\color[HTML]{000000} 0.260} & {\color[HTML]{000000} 0.455} & {\color[HTML]{000000} 0.318} & {\color[HTML]{000000} 0.626} & {\color[HTML]{000000} 0.650} & {\color[HTML]{000000} 0.613} & {\color[HTML]{000000} 0.552} & {\color[HTML]{000000} 0.355} & {\color[HTML]{000000} 0.569} & {\color[HTML]{3166FF} \textbf{0.707}} & {\color[HTML]{FE0000} \textbf{0.711}} \\ \hline
\multicolumn{1}{c|}{{\color[HTML]{000000} }} & {\color[HTML]{000000} PLCC} & {\color[HTML]{000000} 0.084} & {\color[HTML]{000000} 0.193} & {\color[HTML]{000000} 0.415} & {\color[HTML]{000000} 0.178} & {\color[HTML]{000000} 0.765} & {\color[HTML]{000000} 0.710} & {\color[HTML]{000000} 0.733} & {\color[HTML]{000000} 0.716} & {\color[HTML]{000000} 0.595} & {\color[HTML]{000000} 0.626} & {\color[HTML]{3166FF} \textbf{0.855}} & {\color[HTML]{FE0000} \textbf{0.865}} \\
\multicolumn{1}{c|}{{\color[HTML]{000000} }} & {\color[HTML]{000000} SRCC} & {\color[HTML]{000000} 0.038} & {\color[HTML]{000000} 0.150} & {\color[HTML]{000000} 0.527} & {\color[HTML]{000000} 0.254} & {\color[HTML]{000000} 0.785} & {\color[HTML]{000000} 0.694} & {\color[HTML]{000000} 0.714} & {\color[HTML]{000000} 0.696} & {\color[HTML]{000000} 0.508} & {\color[HTML]{000000} 0.542} & {\color[HTML]{3166FF} \textbf{0.846}} & {\color[HTML]{FE0000} \textbf{0.858}} \\
\multicolumn{1}{c|}{{\color[HTML]{000000} }} & {\color[HTML]{000000} RMSE} & {\color[HTML]{000000} 14.413} & {\color[HTML]{000000} 14.337} & {\color[HTML]{000000} 11.382} & {\color[HTML]{000000} 14.055} & {\color[HTML]{000000} 9.324} & {\color[HTML]{000000} 10.219} & {\color[HTML]{3166FF} \textbf{9.331}} & {\color[HTML]{000000} 10.159} & {\color[HTML]{000000} 11.424} & {\color[HTML]{000000} 11.450} & {\color[HTML]{000000} /} & {\color[HTML]{FE0000} \textbf{7.069}} \\
\multicolumn{1}{c|}{\multirow{-4}{*}{{\color[HTML]{000000} WaterlooSQoE-IV}}} & {\color[HTML]{000000} KRCC} & {\color[HTML]{000000} 0.031} & {\color[HTML]{000000} 0.116} & {\color[HTML]{000000} 0.374} & {\color[HTML]{000000} 0.182} & {\color[HTML]{000000} 0.608} & {\color[HTML]{000000} 0.499} & {\color[HTML]{000000} 0.532} & {\color[HTML]{000000} 0.501} & {\color[HTML]{000000} 0.357} & {\color[HTML]{000000} 0.398} & {\color[HTML]{3166FF} \textbf{0.668}} & {\color[HTML]{FE0000} \textbf{0.674}} \\ \hline
\multicolumn{1}{c|}{{\color[HTML]{000000} }} & {\color[HTML]{000000} PLCC} & {\color[HTML]{000000} 0.612} & {\color[HTML]{000000} 0.734} & {\color[HTML]{000000} 0.575} & {\color[HTML]{000000} 0.779} & {\color[HTML]{000000} 0.910} & {\color[HTML]{000000} 0.814} & {\color[HTML]{000000} 0.842} & {\color[HTML]{000000} 0.876} & {\color[HTML]{000000} 0.758} & {\color[HTML]{3166FF} \textbf{0.915}} & {\color[HTML]{000000} /} & {\color[HTML]{FE0000} \textbf{0.959}} \\
\multicolumn{1}{c|}{{\color[HTML]{000000} }} & {\color[HTML]{000000} SRCC} & {\color[HTML]{000000} 0.535} & {\color[HTML]{000000} 0.660} & {\color[HTML]{000000} 0.578} & {\color[HTML]{000000} 0.767} & {\color[HTML]{000000} 0.868} & {\color[HTML]{000000} 0.837} & {\color[HTML]{000000} 0.870} & {\color[HTML]{000000} 0.858} & {\color[HTML]{000000} 0.709} & {\color[HTML]{3166FF} \textbf{0.891}} & {\color[HTML]{000000} /} & {\color[HTML]{FE0000} \textbf{0.925}} \\
\multicolumn{1}{c|}{{\color[HTML]{000000} }} & {\color[HTML]{000000} RMSE} & {\color[HTML]{000000} 0.612} & {\color[HTML]{000000} 0.555} & {\color[HTML]{000000} 0.684} & {\color[HTML]{000000} 0.501} & {\color[HTML]{000000} 0.299} & {\color[HTML]{000000} 0.324} & {\color[HTML]{000000} 0.298} & {\color[HTML]{000000} 0.345} & {\color[HTML]{000000} 0.530} & {\color[HTML]{3166FF} \textbf{0.301}} & {\color[HTML]{000000} /} & {\color[HTML]{FE0000} \textbf{0.230}} \\
\multicolumn{1}{c|}{\multirow{-4}{*}{{\color[HTML]{000000} TaoLive QoE Database}}} & {\color[HTML]{000000} KRCC} & {\color[HTML]{000000} 0.470} & {\color[HTML]{000000} 0.541} & {\color[HTML]{000000} 0.449} & {\color[HTML]{000000} 0.571} & {\color[HTML]{000000} 0.695} & {\color[HTML]{000000} 0.665} & {\color[HTML]{000000} 0.702} & {\color[HTML]{000000} 0.678} & {\color[HTML]{000000} 0.558} & {\color[HTML]{3166FF} \textbf{0.680}} & {\color[HTML]{000000} /} & {\color[HTML]{FE0000} \textbf{0.769}} \\ \hline
\multicolumn{1}{c|}{{\color[HTML]{000000} }} & {\color[HTML]{000000} PLCC} & {\color[HTML]{000000} 0.380} & {\color[HTML]{000000} 0.445} & {\color[HTML]{000000} 0.596} & {\color[HTML]{000000} 0.567} & {\color[HTML]{000000} 0.741} & {\color[HTML]{3166FF} \textbf{0.840}} & {\color[HTML]{000000} 0.811} & {\color[HTML]{000000} 0.784} & {\color[HTML]{000000} 0.750} & {\color[HTML]{000000} 0.767} & {\color[HTML]{000000} /} & {\color[HTML]{FE0000} \textbf{0.913}} \\
\multicolumn{1}{c|}{{\color[HTML]{000000} }} & {\color[HTML]{000000} SRCC} & {\color[HTML]{000000} 0.344} & {\color[HTML]{000000} 0.404} & {\color[HTML]{000000} 0.638} & {\color[HTML]{000000} 0.589} & {\color[HTML]{000000} 0.782} & {\color[HTML]{3166FF} \textbf{0.833}} & {\color[HTML]{000000} 0.807} & {\color[HTML]{000000} 0.785} & {\color[HTML]{000000} 0.669} & {\color[HTML]{000000} 0.728} & {\color[HTML]{000000} /} & {\color[HTML]{FE0000} \textbf{0.902}} \\
\multicolumn{1}{c|}{{\color[HTML]{000000} }} & {\color[HTML]{000000} RMSE} & {\color[HTML]{000000} 10.104} & {\color[HTML]{000000} 10.008} & {\color[HTML]{000000} 7.778} & {\color[HTML]{000000} 8.573} & {\color[HTML]{000000} 6.971} & {\color[HTML]{3166FF} \textbf{5.550}} & {\color[HTML]{000000} 5.952} & {\color[HTML]{000000} 7.141} & {\color[HTML]{000000} 6.788} & {\color[HTML]{000000} 7.092} & {\color[HTML]{000000} /} & {\color[HTML]{FE0000} \textbf{4.642}} \\
\multicolumn{1}{c|}{\multirow{-4}{*}{{\color[HTML]{000000} W.A.}}} & {\color[HTML]{000000} KRCC} & {\color[HTML]{000000} 0.279} & {\color[HTML]{000000} 0.310} & {\color[HTML]{000000} 0.474} & {\color[HTML]{000000} 0.438} & {\color[HTML]{000000} 0.606} & {\color[HTML]{3166FF} \textbf{0.658}} & {\color[HTML]{000000} 0.624} & {\color[HTML]{000000} 0.604} & {\color[HTML]{000000} 0.516} & {\color[HTML]{000000} 0.545} & {\color[HTML]{000000} /} & {\color[HTML]{FE0000} \textbf{0.735}} \\ \hline
\end{tabular}%
}
\vspace{-15pt}
\end{table*}

\begin{figure*}[t]
    \centering
    \subfloat[]{\includegraphics[width=1.3in]{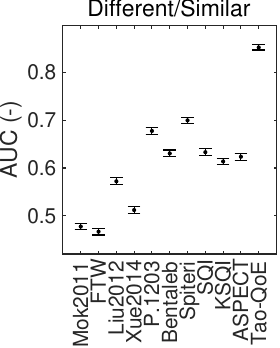}%
    \label{MOS waterloo4_different_similar_AUC}}
    \hfil
    \subfloat[]{\includegraphics[width=1.48in]{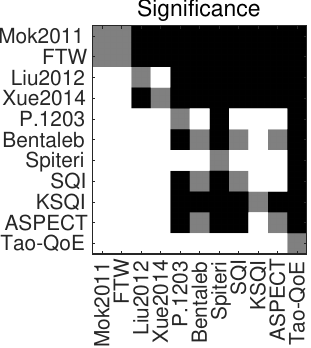}%
    \label{waterloo4_different_similar_significance}}
    \hfil
    \subfloat[]{\includegraphics[width=1.3in]{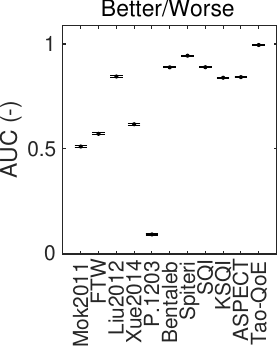}%
    \label{waterloo4_better_worse_AUC}}
    \hfil
    \subfloat[]{\includegraphics[width=1.48in]{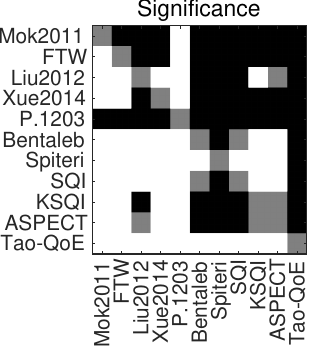}%
    \label{waterloo4_better_worse_significance}}

    \subfloat[]{\includegraphics[width=1.3in]{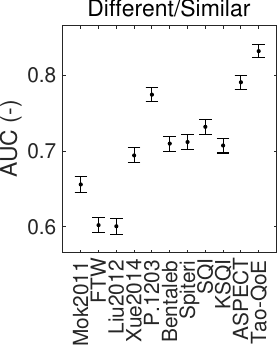}%
    \label{MOS TLQD_different_similar_AUC}}
    \hfil
    \subfloat[]{\includegraphics[width=1.48in]{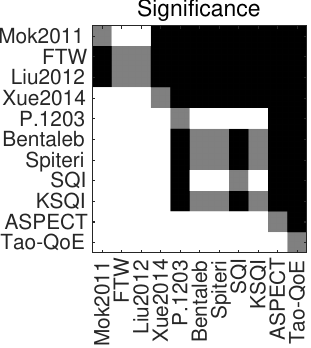}%
    \label{TLQD_different_similar_significance}}
    \hfil
    \subfloat[]{\includegraphics[width=1.3in]{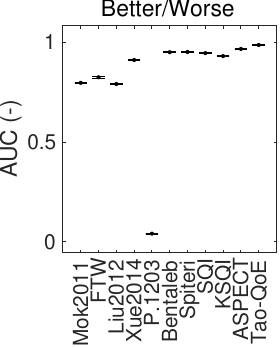}%
    \label{TLQD_better_worse_AUC}}
    \hfil
    \subfloat[]{\includegraphics[width=1.48in]{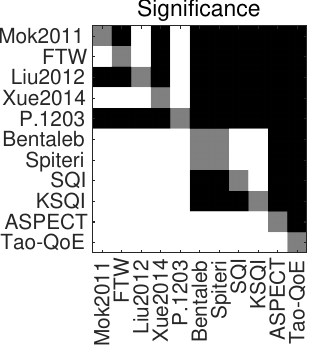}%
    \label{TLQD_better_worse_significance}}

    \caption{New criteria performance of 11 state-of-art FR and NR QoE models and our proposed model on WaterlooSQoE-IV database and TaoLive QoE database. (a), (b) and (e), (f) are the different vs. similar ROC analysis results on WaterlooSQoE-IV database and TaoLive QoE database. (c), (d) and (g), (h) are the better vs. worse analysis results on WaterlooSQoE-IV database and TaoLive QoE database. Note that a white/black square in (b), (d), (f), (h) means the row metric is statistically better/worse than the column one. A gray square means the row method and the column method are statistically indistinguishable.}
    \label{classfication criteria}
    \vspace{-15pt}
\end{figure*}

\subsection{QoE Performance}
\subsubsection{VQEG Criteria}

Table~\ref{tab:QoE compare} presents a performance comparison between our proposed Tao-QoE model and existing mainstream models across six databases. First, the Tao-QoE model achieves the highest PLCC (0.913) and SRCC (0.902) in terms of weighted average (W.A.) metrics, demonstrating its overall superiority. Notably, the most significant performance improvements are observed on databases containing complex authentic distortions. For instance, on the TaoLive QoE Database, which includes real-world impairments from live streaming scenarios, our model substantially outperforms other methods. This advantage arises because Tao-QoE is specifically designed for live video, enabling joint modeling of source encoding distortions and transport-layer impairments (e.g., stalling events and accelerated playing), both of which are critical factors affecting user experience in live streaming contexts. In contrast, on databases dominated by traditional spatial distortions (e.g., LIVE-II), the performance gaps among state-of-the-art models are relatively small, indicating that the field of pure spatial quality assessment has matured, leaving limited room for further improvement. Second, the performance gap between deep learning-based models and traditional approaches is particularly pronounced on databases containing quality switches and stalling events (WaterlooSQoE-III/IV). Traditional models (e.g., Mok2011, FTW) rely on handcrafted features derived from network-layer parameters and assume a linear relationship between these features and user experience. However, human perception of quality fluctuations and stalling exhibits nonlinearity and memory effects, which cannot be captured by the linear assumptions of traditional models. In contrast, deep learning models can automatically learn hierarchical features from video data, effectively modeling such nonlinear temporal dependencies. Consequently, these models demonstrate significant advantages in complex distortion scenarios.


\subsubsection{Classfication Criteria}
We present the performance evaluation based on the classification criteria of all the QoE models using TaoLive QoE database and the largest publicly available database, WaterlooSQoE-IV, as shown in Fig \ref{classfication criteria}. From this figure, we can draw similar conclusions to those derived from the VQEG performance. Firstly, our proposed model Tao-QoE outperforms other QoE models by a significant margin in both the \emph{'Different vs. Similar'} and \emph{'Better vs. Worse'} classification tasks. Statistical analysis also demonstrates that our proposed model is significantly superior to other models on the TaoLive QoE database and WaterlooSQoE-IV database. Secondly, the AUC values for the \emph{'Better vs. Worse'} classification task are consistently higher than those for the \emph{'Different vs. Similar'} classification task, indicating that the latter is more challenging and there is still room for improvement in this area.


\begin{table}[ht]
\centering
\caption{Comparison of VQA models. Best in {\color[HTML]{FE0000} \textbf{red}} and second in {\color[HTML]{3166FF} \textbf{blue}}}
\label{tab:VQA compare}
\adjustbox{max width=\columnwidth}{
\begin{tabular}{cc|c|c|c|c|c}
\hline
\multicolumn{2}{c|}{Models} &  &  &  &  &  \\ \cline{1-2}
\multicolumn{1}{c|}{Databases} & Criteria & \multirow{-2}{*}{TLVQM} & \multirow{-2}{*}{VSFA} & \multirow{-2}{*}{SimpleVQA} & \multirow{-2}{*}{FastVQA} & \multirow{-2}{*}{Tao-QoE} \\ \hline
\multicolumn{1}{c|}{} & PLCC & 0.625 & 0.461 & {\color[HTML]{3166FF} \textbf{0.783}} & 0.584 & {\color[HTML]{FE0000} \textbf{0.802}} \\
\multicolumn{1}{c|}{} & SRCC & 0.599 & 0.458 & {\color[HTML]{3166FF} \textbf{0.758}} & 0.547 & {\color[HTML]{FE0000} \textbf{0.768}} \\
\multicolumn{1}{c|}{} & RMSE & 9.017 & 7.293 & {\color[HTML]{3166FF} \textbf{7.100}} & 9.014 & {\color[HTML]{FE0000} \textbf{6.848}} \\
\multicolumn{1}{c|}{\multirow{-4}{*}{LIVE-Qualcomm}} & KRCC & 0.453 & 0.330 & {\color[HTML]{3166FF} \textbf{0.573}} & 0.385 & {\color[HTML]{FE0000} \textbf{0.580}} \\ \hline
\multicolumn{1}{c|}{} & PLCC & 0.771 & 0.449 & {\color[HTML]{3166FF} \textbf{0.890}} & 0.850 & {\color[HTML]{FE0000} \textbf{0.906}} \\
\multicolumn{1}{c|}{} & SRCC & 0.751 & 0.347 & {\color[HTML]{3166FF} \textbf{0.875}} & 0.843 & {\color[HTML]{FE0000} \textbf{0.898}} \\
\multicolumn{1}{c|}{} & RMSE & 13.925 & 18.548 & {\color[HTML]{3166FF} \textbf{9.694}} & 12.619 & {\color[HTML]{FE0000} \textbf{9.038}} \\
\multicolumn{1}{c|}{\multirow{-4}{*}{CVD2014}} & KRCC & 0.565 & 0.246 & {\color[HTML]{3166FF} \textbf{0.703}} & 0.651 & {\color[HTML]{FE0000} \textbf{0.728}} \\ \hline
\multicolumn{1}{c|}{} & PLCC & 0.843 & 0.774 & {\color[HTML]{3166FF} \textbf{0.845}} & 0.759 & {\color[HTML]{FE0000} \textbf{0.868}} \\
\multicolumn{1}{c|}{} & SRCC & 0.810 & 0.768 & {\color[HTML]{3166FF} \textbf{0.842}} & 0.759 & {\color[HTML]{FE0000} \textbf{0.867}} \\
\multicolumn{1}{c|}{} & RMSE & 0.355 & 0.413 & {\color[HTML]{3166FF} \textbf{0.349}} & 0.459 & {\color[HTML]{FE0000} \textbf{0.324}} \\
\multicolumn{1}{c|}{\multirow{-4}{*}{KoNViD-1k}} & KRCC & 0.615 & 0.572 & {\color[HTML]{3166FF} \textbf{0.651}} & 0.569 & {\color[HTML]{FE0000} \textbf{0.679}} \\ \hline
\multicolumn{1}{c|}{} & PLCC & 0.577 & 0.534 & {\color[HTML]{3166FF} \textbf{0.726}} & 0.592 & {\color[HTML]{FE0000} \textbf{0.749}} \\
\multicolumn{1}{c|}{} & SRCC & 0.573 & 0.513 & {\color[HTML]{3166FF} \textbf{0.724}} & 0.622 & {\color[HTML]{FE0000} \textbf{0.753}} \\
\multicolumn{1}{c|}{} & RMSE & 11.321 & 11.127 & {\color[HTML]{3166FF} \textbf{9.277}} & 11.422 & {\color[HTML]{FE0000} \textbf{8.956}} \\
\multicolumn{1}{c|}{\multirow{-4}{*}{VDPVE}} & KRCC & 0.406 & 0.365 & {\color[HTML]{3166FF} \textbf{0.532}} & 0.436 & {\color[HTML]{FE0000} \textbf{0.555}} \\ \hline
\multicolumn{1}{c|}{} & PLCC & {\color[HTML]{3166FF} \textbf{0.789}} & 0.765 & 0.747 & 0.709 & {\color[HTML]{FE0000} \textbf{0.869}} \\
\multicolumn{1}{c|}{} & SRCC & {\color[HTML]{3166FF} \textbf{0.786}} & 0.737 & 0.716 & 0.695 & {\color[HTML]{FE0000} \textbf{0.857}} \\
\multicolumn{1}{c|}{} & RMSE & {\color[HTML]{3166FF} \textbf{10.708}} & 11.061 & 11.471 & 13.222 & {\color[HTML]{FE0000} \textbf{8.537}} \\
\multicolumn{1}{c|}{\multirow{-4}{*}{LIVE-VQC}} & KRCC & {\color[HTML]{3166FF} \textbf{0.595}} & 0.544 & 0.527 & 0.512 & {\color[HTML]{FE0000} \textbf{0.680}} \\ \hline
\multicolumn{1}{c|}{} & PLCC & {\color[HTML]{000000} 0.411} & 0.582 & {\color[HTML]{3166FF} \textbf{0.789}} & 0.631 & {\color[HTML]{FE0000} \textbf{0.805}} \\
\multicolumn{1}{c|}{} & SRCC & 0.391 & 0.528 & {\color[HTML]{3166FF} \textbf{0.767}} & 0.609 & {\color[HTML]{FE0000} \textbf{0.777}} \\
\multicolumn{1}{c|}{} & RMSE & 2.719 & 1.101 & {\color[HTML]{3166FF} \textbf{1.039}} & 1.332 & {\color[HTML]{FE0000} \textbf{1.010}} \\
\multicolumn{1}{c|}{\multirow{-4}{*}{MSU}} & KRCC & 0.280 & 0.399 & {\color[HTML]{3166FF} \textbf{0.589}} & 0.438 & {\color[HTML]{FE0000} \textbf{0.584}} \\ \hline
\multicolumn{1}{c|}{} & PLCC & 0.624 & 0.513 & {\color[HTML]{3166FF} \textbf{0.812}} & 0.535 & {\color[HTML]{FE0000} \textbf{0.854}} \\
\multicolumn{1}{c|}{} & SRCC & 0.662 & 0.532 & {\color[HTML]{3166FF} \textbf{0.815}} & 0.536 & {\color[HTML]{FE0000} \textbf{0.852}} \\
\multicolumn{1}{c|}{} & RMSE & 0.537 & 0.535 & {\color[HTML]{3166FF} \textbf{0.381}} & 0.654 & {\color[HTML]{FE0000} \textbf{0.340}} \\
\multicolumn{1}{c|}{\multirow{-4}{*}{YouTubeUGC}} & KRCC & 0.472 & 0.383 & {\color[HTML]{3166FF} \textbf{0.621}} & 0.370 & {\color[HTML]{FE0000} \textbf{0.662}} \\ \hline
\multicolumn{1}{c|}{} & PLCC & 0.823 & 0.589 & {\color[HTML]{3166FF} \textbf{0.922}} & 0.764 & {\color[HTML]{FE0000} \textbf{0.945}} \\
\multicolumn{1}{c|}{} & SRCC & 0.824 & 0.597 & {\color[HTML]{3166FF} \textbf{0.921}} & 0.773 & {\color[HTML]{FE0000} \textbf{0.942}} \\
\multicolumn{1}{c|}{} & RMSE & 8.998 & 10.517 & {\color[HTML]{3166FF} \textbf{5.537}} & 9.203 & {\color[HTML]{FE0000} \textbf{4.706}} \\
\multicolumn{1}{c|}{\multirow{-4}{*}{LIVE-WC}} & KRCC & 0.638 & 0.463 & {\color[HTML]{3166FF} \textbf{0.756}} & 0.575 & {\color[HTML]{FE0000} \textbf{0.794}} \\ \hline
\multicolumn{1}{c|}{} & PLCC & 0.688 & 0.454 & {\color[HTML]{3166FF} \textbf{0.750}} & 0.584 & {\color[HTML]{FE0000} \textbf{0.914}} \\
\multicolumn{1}{c|}{} & SRCC & 0.645 & 0.415 & {\color[HTML]{3166FF} \textbf{0.714}} & 0.600 & {\color[HTML]{FE0000} \textbf{0.908}} \\
\multicolumn{1}{c|}{} & RMSE & 7.745 & 11.236 & {\color[HTML]{3166FF} \textbf{8.721}} & 11.574 & {\color[HTML]{FE0000} \textbf{5.169}} \\
\multicolumn{1}{c|}{\multirow{-4}{*}{LIVE-APV}} & KRCC & 0.547 & 0.283 & {\color[HTML]{3166FF} \textbf{0.530}} & 0.422 & {\color[HTML]{FE0000} \textbf{0.717}} \\ \hline
\multicolumn{1}{c|}{} & PLCC & 0.690 & 0.568 & {\color[HTML]{3166FF} \textbf{0.807}} & 0.670 & {\color[HTML]{FE0000} \textbf{0.857}} \\
\multicolumn{1}{c|}{} & SRCC & 0.683 & 0.543 & {\color[HTML]{3166FF} \textbf{0.796}} & 0.639 & {\color[HTML]{FE0000} \textbf{0.839}} \\
\multicolumn{1}{c|}{} & RMSE & 7.369 & 8.081 & {\color[HTML]{3166FF} \textbf{5.891}} & 7.838 & {\color[HTML]{FE0000} \textbf{4.933}} \\
\multicolumn{1}{c|}{\multirow{-4}{*}{W.A.}} & KRCC & 0.508 & 0.390 & {\color[HTML]{3166FF} \textbf{0.621}} & 0.484 & {\color[HTML]{FE0000} \textbf{0.667}} \\ \hline
\end{tabular}
}
\vspace{-10pt}
\end{table}

\subsection{VQA Performance}

We present the VQEG performance on 9 UGC VQA databases in Table \ref{tab:VQA compare}. First, the proposed Tao-QoE model achieves the best performance among all compared methods, with particularly significant improvements on three large-scale UGC databases: MSU, YouTubeUGC, and LIVE-WC. For instance, on the LIVE-WC database, Tao-QoE attains a PLCC of 0.945, outperforming the suboptimal model SimpleVQA (0.922) by a margin of 2.3\%. This advantage is primarily attributed to the complex mixed distortions present in UGC videos (e.g., camera shake, overexposure, blur). Through large-scale pre-training, Tao-QoE acquires robust representations of general distortion types, enabling it to effectively handle the complexity of UGC content and demonstrate strong cross-scenario generalization capability. Second, deep learning-based models exhibit clear advantages on UGC data. Traditional methods represented by TLVQM rely on handcrafted features and are constrained by prior knowledge, making it difficult to cover the diverse and emerging distortion patterns in UGC videos. In contrast, deep learning can automatically learn hierarchical representations of distortions, offering greater adaptability. Third, although live streaming and UGC videos differ in application scenarios, they share common challenges—uncontrollable source content quality and diverse distortion types. This provides insights for developing a unified cross-scenario quality assessment model: the two tasks can mutually reinforce each other, jointly benefiting from deep modeling of complex authentic distortions. Finally, from the weighted average metrics, Tao-QoE achieves leading performance in PLCC (0.857), SRCC (0.839), and the lowest RMSE (4.933), demonstrating its consistency and stability across different UGC databases. Such stability is crucial in practical applications, as real-world video quality assessment systems must handle video content from diverse sources with unknown distortion types. The robustness of the model directly determines its practical utility.

\begin{table}[ht]
\centering
\caption{Experimental performance of the ablation study of QoE databases. Best in {\color[HTML]{FE0000} \textbf{red}} and second in {\color[HTML]{3166FF} \textbf{blue}}. S, F, FM, M represent semantic feature extraction module, multi-scale feature fusion module, optiacl flow motion feature extraction module, motion feature extraction module respectively. ALL represents S+FM+F.}
\label{tab:ablation of QoE databases}
\resizebox{\columnwidth}{!}{%
\begin{tabular}{cc|c|c|c|c|c|c}
\hline
\multicolumn{2}{c|}{{\color[HTML]{000000} Models}} & {\color[HTML]{000000} } & {\color[HTML]{000000} } & {\color[HTML]{000000} } & {\color[HTML]{000000} } & {\color[HTML]{000000} } & {\color[HTML]{000000} } \\ \cline{1-2}
\multicolumn{1}{c|}{{\color[HTML]{000000} Databases}} & {\color[HTML]{000000} Criteria} & \multirow{-2}{*}{{\color[HTML]{000000} S}} & \multirow{-2}{*}{{\color[HTML]{000000} S+F}} & \multirow{-2}{*}{{\color[HTML]{000000} FM}} & \multirow{-2}{*}{{\color[HTML]{000000} S+FM}} & \multirow{-2}{*}{{\color[HTML]{000000} S+F+M}} & \multirow{-2}{*}{{\color[HTML]{000000} ALL}} \\ \hline
\multicolumn{1}{c|}{{\color[HTML]{000000} }} & {\color[HTML]{000000} PLCC} & {\color[HTML]{3166FF} \textbf{0.912}} & {\color[HTML]{000000} 0.773} & {\color[HTML]{000000} 0.858} & {\color[HTML]{000000} 0.909} & {\color[HTML]{000000} 0.835} & {\color[HTML]{FE0000} \textbf{0.948}} \\
\multicolumn{1}{c|}{{\color[HTML]{000000} }} & {\color[HTML]{000000} SRCC} & {\color[HTML]{3166FF} \textbf{0.904}} & {\color[HTML]{000000} 0.778} & {\color[HTML]{000000} 0.832} & {\color[HTML]{000000} 0.908} & {\color[HTML]{000000} 0.810} & {\color[HTML]{FE0000} \textbf{0.946}} \\
\multicolumn{1}{c|}{{\color[HTML]{000000} }} & {\color[HTML]{000000} RMSE} & {\color[HTML]{3166FF} \textbf{0.304}} & {\color[HTML]{000000} 0.441} & {\color[HTML]{000000} 0.394} & {\color[HTML]{000000} 0.323} & {\color[HTML]{000000} 0.420} & {\color[HTML]{FE0000} \textbf{0.250}} \\
\multicolumn{1}{c|}{\multirow{-4}{*}{{\color[HTML]{000000} LIVE-NFLX-II}}} & {\color[HTML]{000000} KRCC} & {\color[HTML]{3166FF} \textbf{0.746}} & {\color[HTML]{000000} 0.618} & {\color[HTML]{000000} 0.659} & {\color[HTML]{000000} 0.743} & {\color[HTML]{000000} 0.651} & {\color[HTML]{FE0000} \textbf{0.800}} \\ \hline
\multicolumn{1}{c|}{{\color[HTML]{000000} }} & {\color[HTML]{000000} PLCC} & {\color[HTML]{000000} 0.874} & {\color[HTML]{000000} 0.907} & {\color[HTML]{000000} 0.704} & {\color[HTML]{3166FF} \textbf{0.925}} & {\color[HTML]{000000} 0.900} & {\color[HTML]{FE0000} \textbf{0.933}} \\
\multicolumn{1}{c|}{{\color[HTML]{000000} }} & {\color[HTML]{000000} SRCC} & {\color[HTML]{000000} 0.869} & {\color[HTML]{000000} 0.909} & {\color[HTML]{000000} 0.686} & {\color[HTML]{3166FF} \textbf{0.926}} & {\color[HTML]{000000} 0.891} & {\color[HTML]{FE0000} \textbf{0.929}} \\
\multicolumn{1}{c|}{{\color[HTML]{000000} }} & {\color[HTML]{000000} RMSE} & {\color[HTML]{000000} 9.622} & {\color[HTML]{000000} 8.353} & {\color[HTML]{000000} 13.861} & {\color[HTML]{3166FF} \textbf{7.514}} & {\color[HTML]{000000} 8.438} & {\color[HTML]{FE0000} \textbf{7.163}} \\
\multicolumn{1}{c|}{\multirow{-4}{*}{{\color[HTML]{000000} WaterlooSQoE-I}}} & {\color[HTML]{000000} KRCC} & {\color[HTML]{000000} 0.688} & {\color[HTML]{000000} 0.743} & {\color[HTML]{000000} 0.515} & {\color[HTML]{3166FF} \textbf{0.777}} & {\color[HTML]{000000} 0.724} & {\color[HTML]{FE0000} \textbf{0.775}} \\ \hline
\multicolumn{1}{c|}{{\color[HTML]{000000} }} & {\color[HTML]{000000} PLCC} & {\color[HTML]{000000} 0.800} & {\color[HTML]{3166FF} \textbf{0.868}} & {\color[HTML]{000000} 0.762} & {\color[HTML]{000000} 0.809} & {\color[HTML]{000000} 0.814} & {\color[HTML]{FE0000} \textbf{0.874}} \\
\multicolumn{1}{c|}{{\color[HTML]{000000} }} & {\color[HTML]{000000} SRCC} & {\color[HTML]{000000} 0.797} & {\color[HTML]{3166FF} \textbf{0.861}} & {\color[HTML]{000000} 0.730} & {\color[HTML]{000000} 0.796} & {\color[HTML]{000000} 0.793} & {\color[HTML]{FE0000} \textbf{0.866}} \\
\multicolumn{1}{c|}{{\color[HTML]{000000} }} & {\color[HTML]{000000} RMSE} & {\color[HTML]{000000} 7.867} & {\color[HTML]{3166FF} \textbf{6.716}} & {\color[HTML]{000000} 8.647} & {\color[HTML]{000000} 7.779} & {\color[HTML]{000000} 7.908} & {\color[HTML]{FE0000} \textbf{6.645}} \\
\multicolumn{1}{c|}{\multirow{-4}{*}{{\color[HTML]{000000} WaterlooSQoE-II}}} & {\color[HTML]{000000} KRCC} & {\color[HTML]{000000} 0.614} & {\color[HTML]{3166FF} \textbf{0.678}} & {\color[HTML]{000000} 0.554} & {\color[HTML]{000000} 0.607} & {\color[HTML]{000000} 0.604} & {\color[HTML]{FE0000} \textbf{0.680}} \\ \hline
\multicolumn{1}{c|}{{\color[HTML]{000000} }} & {\color[HTML]{000000} PLCC} & {\color[HTML]{000000} 0.751} & {\color[HTML]{3166FF} \textbf{0.899}} & {\color[HTML]{000000} 0.642} & {\color[HTML]{000000} 0.867} & {\color[HTML]{000000} 0.886} & {\color[HTML]{FE0000} \textbf{0.900}} \\
\multicolumn{1}{c|}{{\color[HTML]{000000} }} & {\color[HTML]{000000} SRCC} & {\color[HTML]{000000} 0.628} & {\color[HTML]{3166FF} \textbf{0.884}} & {\color[HTML]{000000} 0.511} & {\color[HTML]{000000} 0.847} & {\color[HTML]{000000} 0.873} & {\color[HTML]{FE0000} \textbf{0.890}} \\
\multicolumn{1}{c|}{{\color[HTML]{000000} }} & {\color[HTML]{000000} RMSE} & {\color[HTML]{000000} 9.697} & {\color[HTML]{3166FF} \textbf{6.529}} & {\color[HTML]{000000} 11.329} & {\color[HTML]{000000} 7.386} & {\color[HTML]{000000} 6.837} & {\color[HTML]{FE0000} \textbf{6.494}} \\
\multicolumn{1}{c|}{\multirow{-4}{*}{{\color[HTML]{000000} WaterlooSQoE-III}}} & {\color[HTML]{000000} KRCC} & {\color[HTML]{000000} 0.468} & {\color[HTML]{3166FF} \textbf{0.706}} & {\color[HTML]{000000} 0.372} & {\color[HTML]{000000} 0.658} & {\color[HTML]{000000} 0.695} & {\color[HTML]{FE0000} \textbf{0.711}} \\ \hline
\multicolumn{1}{c|}{{\color[HTML]{000000} }} & {\color[HTML]{000000} PLCC} & {\color[HTML]{000000} 0.814} & {\color[HTML]{3166FF} \textbf{0.855}} & {\color[HTML]{000000} 0.806} & {\color[HTML]{000000} 0.843} & {\color[HTML]{000000} 0.847} & {\color[HTML]{FE0000} \textbf{0.865}} \\
\multicolumn{1}{c|}{{\color[HTML]{000000} }} & {\color[HTML]{000000} SRCC} & {\color[HTML]{000000} 0.791} & {\color[HTML]{3166FF} \textbf{0.847}} & {\color[HTML]{000000} 0.779} & {\color[HTML]{000000} 0.831} & {\color[HTML]{000000} 0.835} & {\color[HTML]{FE0000} \textbf{0.858}} \\
\multicolumn{1}{c|}{{\color[HTML]{000000} }} & {\color[HTML]{000000} RMSE} & {\color[HTML]{000000} 8.221} & {\color[HTML]{3166FF} \textbf{7.241}} & {\color[HTML]{000000} 8.366} & {\color[HTML]{000000} 7.523} & {\color[HTML]{000000} 7.362} & {\color[HTML]{FE0000} \textbf{7.069}} \\
\multicolumn{1}{c|}{\multirow{-4}{*}{{\color[HTML]{000000} WaterlooSQoE-IV}}} & {\color[HTML]{000000} KRCC} & {\color[HTML]{000000} 0.604} & {\color[HTML]{3166FF} \textbf{0.664}} & {\color[HTML]{000000} 0.609} & {\color[HTML]{000000} 0.645} & {\color[HTML]{000000} 0.643} & {\color[HTML]{FE0000} \textbf{0.674}} \\ \hline

\multicolumn{1}{c|}{{\color[HTML]{000000} }} & {\color[HTML]{000000} PLCC} & {\color[HTML]{000000} 0.908} & {\color[HTML]{000000} 0.923} & {\color[HTML]{000000} 0.694} & {\color[HTML]{3166FF} \textbf{0.947}} & {\color[HTML]{000000} 0.938} & {\color[HTML]{FE0000} \textbf{0.959}} \\
\multicolumn{1}{c|}{{\color[HTML]{000000} }} & {\color[HTML]{000000} SRCC} & {\color[HTML]{000000} 0.865} & {\color[HTML]{000000} 0.875} & {\color[HTML]{000000} 0.650} & {\color[HTML]{3166FF} \textbf{0.921}} & {\color[HTML]{000000} 0.912} & {\color[HTML]{FE0000} \textbf{0.925}} \\
\multicolumn{1}{c|}{{\color[HTML]{000000} }} & {\color[HTML]{000000} RMSE} & {\color[HTML]{000000} 0.398} & {\color[HTML]{000000} 0.314} & {\color[HTML]{000000} 0.585} & {\color[HTML]{3166FF} \textbf{0.266}} & {\color[HTML]{000000} 0.300} & {\color[HTML]{FE0000} \textbf{0.230}} \\
\multicolumn{1}{c|}{\multirow{-4}{*}{{\color[HTML]{000000} TLVD}}} & {\color[HTML]{000000} KRCC} & {\color[HTML]{000000} 0.643} & {\color[HTML]{000000} 0.702} & {\color[HTML]{000000} 0.473} & {\color[HTML]{3166FF} \textbf{0.764}} & {\color[HTML]{000000} 0.755} & {\color[HTML]{FE0000} \textbf{0.769}} \\ \hline
\end{tabular}%
}
\vspace{-15pt}
\end{table}

\begin{table}[ht]
\centering
\caption{Experimental performance of the ablation study of VQA databases. Best in {\color[HTML]{FE0000} \textbf{red}} and second in {\color[HTML]{3166FF} \textbf{blue}}. S, F, FM, M represent semantic feature extraction module, multi-scale feature fusion module, optiacl flow motion feature extraction module, motion feature extraction module respectively. ALL represents S+FM+F.}
\label{tab:ablation of VQA databases}
\resizebox{\columnwidth}{!}{%
\begin{tabular}{cc|c|c|c|c|l|c}
\hline
\multicolumn{2}{c|}{{\color[HTML]{000000} Models}} & {\color[HTML]{000000} } & {\color[HTML]{000000} } & {\color[HTML]{000000} } & {\color[HTML]{000000} } & {\color[HTML]{000000} } & {\color[HTML]{000000} } \\ \cline{1-2}
\multicolumn{1}{c|}{{\color[HTML]{000000} Databases}} & {\color[HTML]{000000} Criteria} & \multirow{-2}{*}{{\color[HTML]{000000} S}} & \multirow{-2}{*}{{\color[HTML]{000000} S+F}} & \multirow{-2}{*}{{\color[HTML]{000000} FM}} & \multirow{-2}{*}{{\color[HTML]{000000} S+FM}} & \multirow{-2}{*}{{\color[HTML]{000000} S+F+M}} & \multirow{-2}{*}{{\color[HTML]{000000} ALL}} \\ \hline
\multicolumn{1}{c|}{{\color[HTML]{000000} }} & {\color[HTML]{000000} PLCC} & {\color[HTML]{000000} 0.776} & {\color[HTML]{000000} 0.785} & {\color[HTML]{000000} 0.621} & {\color[HTML]{FE0000} \textbf{0.808}} & {\color[HTML]{000000} 0.728} & {\color[HTML]{3166FF} \textbf{0.802}} \\

\multicolumn{1}{c|}{{\color[HTML]{000000} }} & {\color[HTML]{000000} SRCC} & {\color[HTML]{000000} 0.730} & {\color[HTML]{000000} 0.737} & {\color[HTML]{000000} 0.525} & {\color[HTML]{FE0000} \textbf{0.777}} & {\color[HTML]{000000} 0.678} & {\color[HTML]{3166FF} \textbf{0.768}} \\
\multicolumn{1}{c|}{{\color[HTML]{000000} }} & {\color[HTML]{000000} RMSE} & {\color[HTML]{000000} 7.183} & {\color[HTML]{000000} 7.057} & {\color[HTML]{000000} 8.895} & {\color[HTML]{FE0000} \textbf{6.721}} & {\color[HTML]{000000} 7.778} & {\color[HTML]{3166FF} \textbf{6.848}} \\
\multicolumn{1}{c|}{\multirow{-4}{*}{{\color[HTML]{000000} LIVE-Qualcomm}}} & {\color[HTML]{000000} KRCC} & {\color[HTML]{000000} 0.553} & {\color[HTML]{000000} 0.556} & {\color[HTML]{000000} 0.378} & {\color[HTML]{FE0000} \textbf{0.584}} & {\color[HTML]{000000} 0.506} & {\color[HTML]{3166FF} \textbf{0.580}} \\ \hline
\multicolumn{1}{c|}{{\color[HTML]{000000} }} & {\color[HTML]{000000} PLCC} & {\color[HTML]{000000} 0.856} & {\color[HTML]{000000} 0.877} & {\color[HTML]{000000} 0.774} & {\color[HTML]{3166FF} \textbf{0.895}} & {\color[HTML]{000000} 0.867} & {\color[HTML]{FE0000} \textbf{0.906}} \\
\multicolumn{1}{c|}{{\color[HTML]{000000} }} & {\color[HTML]{000000} SRCC} & {\color[HTML]{000000} 0.839} & {\color[HTML]{000000} 0.860} & {\color[HTML]{000000} 0.745} & {\color[HTML]{3166FF} \textbf{0.885}} & {\color[HTML]{000000} 0.852} & {\color[HTML]{FE0000} \textbf{0.898}} \\
\multicolumn{1}{c|}{{\color[HTML]{000000} }} & {\color[HTML]{000000} RMSE} & {\color[HTML]{000000} 10.871} & {\color[HTML]{000000} 10.100} & {\color[HTML]{000000} 12.943} & {\color[HTML]{3166FF} \textbf{9.495}} & {\color[HTML]{000000} 10.448} & {\color[HTML]{FE0000} \textbf{9.038}} \\
\multicolumn{1}{c|}{\multirow{-4}{*}{{\color[HTML]{000000} CVD2014}}} & {\color[HTML]{000000} KRCC} & {\color[HTML]{000000} 0.656} & {\color[HTML]{000000} 0.681} & {\color[HTML]{000000} 0.550} & {\color[HTML]{3166FF} \textbf{0.704}} & {\color[HTML]{000000} 0.668} & {\color[HTML]{FE0000} \textbf{0.728}} \\ \hline

\multicolumn{1}{c|}{{\color[HTML]{000000} }} & {\color[HTML]{000000} PLCC} & {\color[HTML]{000000} 0.850} & {\color[HTML]{000000} 0.853} & {\color[HTML]{000000} 0.760} & {\color[HTML]{3166FF} \textbf{0.868}} & {\color[HTML]{000000} 0.857} & {\color[HTML]{FE0000} \textbf{0.874}} \\
\multicolumn{1}{c|}{{\color[HTML]{000000} }} & {\color[HTML]{000000} SRCC} & {\color[HTML]{000000} 0.846} & {\color[HTML]{000000} 0.850} & {\color[HTML]{000000} 0.738} & {\color[HTML]{3166FF} \textbf{0.867}} & {\color[HTML]{000000} 0.851} & {\color[HTML]{FE0000} \textbf{0.871}} \\
\multicolumn{1}{c|}{{\color[HTML]{000000} }} & {\color[HTML]{000000} RMSE} & {\color[HTML]{000000} 0.341} & {\color[HTML]{000000} 0.340} & {\color[HTML]{000000} 0.428} & {\color[HTML]{3166FF} \textbf{0.324}} & {\color[HTML]{000000} 0.332} & {\color[HTML]{FE0000} \textbf{0.318}} \\
\multicolumn{1}{c|}{\multirow{-4}{*}{{\color[HTML]{000000} KoNViD-1k}}} & {\color[HTML]{000000} KRCC} & {\color[HTML]{000000} 0.654} & {\color[HTML]{000000} 0.658} & {\color[HTML]{000000} 0.547} & {\color[HTML]{3166FF} \textbf{0.679}} & {\color[HTML]{000000} 0.660} & {\color[HTML]{FE0000} \textbf{0.684}} \\ \hline

\multicolumn{1}{c|}{{\color[HTML]{000000} }} & {\color[HTML]{000000} PLCC} & {\color[HTML]{000000} 0.749} & {\color[HTML]{3166FF} \textbf{0.775}} & {\color[HTML]{000000} 0.552} & {\color[HTML]{FE0000} \textbf{0.776}} & {\color[HTML]{000000} 0.754} & {\color[HTML]{000000} 0.757} \\
\multicolumn{1}{c|}{{\color[HTML]{000000} }} & {\color[HTML]{000000} SRCC} & {\color[HTML]{000000} 0.753} & {\color[HTML]{3166FF} \textbf{0.768}} & {\color[HTML]{000000} 0.545} & {\color[HTML]{FE0000} \textbf{0.769}} & {\color[HTML]{000000} 0.754} & {\color[HTML]{000000} 0.759} \\
\multicolumn{1}{c|}{{\color[HTML]{000000} }} & {\color[HTML]{000000} RMSE} & {\color[HTML]{000000} 8.956} & {\color[HTML]{3166FF} \textbf{8.825}} & {\color[HTML]{000000} 11.285} & {\color[HTML]{FE0000} \textbf{8.822}} & {\color[HTML]{000000} 9.025} & {\color[HTML]{000000} 8.961} \\
\multicolumn{1}{c|}{\multirow{-4}{*}{{\color[HTML]{000000} VDPVE}}} & {\color[HTML]{000000} KRCC} & {\color[HTML]{000000} 0.555} & {\color[HTML]{3166FF} \textbf{0.574}} & {\color[HTML]{000000} 0.376} & {\color[HTML]{FE0000} \textbf{0.575}} & {\color[HTML]{000000} 0.559} & {\color[HTML]{000000} 0.562} \\ \hline
\multicolumn{1}{c|}{{\color[HTML]{000000} }} & {\color[HTML]{000000} PLCC} & {\color[HTML]{000000} 0.792} & {\color[HTML]{000000} 0.821} & {\color[HTML]{000000} 0.746} & {\color[HTML]{3166FF} \textbf{0.867}} & {\color[HTML]{000000} 0.823} & {\color[HTML]{FE0000} \textbf{0.869}} \\
\multicolumn{1}{c|}{{\color[HTML]{000000} }} & {\color[HTML]{000000} SRCC} & {\color[HTML]{000000} 0.751} & {\color[HTML]{000000} 0.792} & {\color[HTML]{000000} 0.718} & {\color[HTML]{3166FF} \textbf{0.852}} & {\color[HTML]{000000} 0.798} & {\color[HTML]{FE0000} \textbf{0.857}} \\
\multicolumn{1}{c|}{{\color[HTML]{000000} }} & {\color[HTML]{000000} RMSE} & {\color[HTML]{000000} 10.464} & {\color[HTML]{000000} 9.821} & {\color[HTML]{000000} 11.469} & {\color[HTML]{3166FF} \textbf{8.597}} & {\color[HTML]{000000} 9.749} & {\color[HTML]{FE0000} \textbf{8.537}} \\
\multicolumn{1}{c|}{\multirow{-4}{*}{{\color[HTML]{000000} LIVE-VQC}}} & {\color[HTML]{000000} KRCC} & {\color[HTML]{000000} 0.558} & {\color[HTML]{000000} 0.600} & {\color[HTML]{000000} 0.538} & {\color[HTML]{3166FF} \textbf{0.673}} & {\color[HTML]{000000} 0.607} & {\color[HTML]{FE0000} \textbf{0.680}} \\ \hline
\multicolumn{1}{c|}{{\color[HTML]{000000} }} & {\color[HTML]{000000} PLCC} & {\color[HTML]{000000} 0.696} & {\color[HTML]{000000} 0.766} & {\color[HTML]{000000} 0.647} & {\color[HTML]{3166FF} \textbf{0.784}} & {\color[HTML]{000000} 0.725} & {\color[HTML]{FE0000} \textbf{0.805}} \\
\multicolumn{1}{c|}{{\color[HTML]{000000} }} & {\color[HTML]{000000} SRCC} & {\color[HTML]{000000} 0.660} & {\color[HTML]{000000} 0.722} & {\color[HTML]{000000} 0.573} & {\color[HTML]{3166FF} \textbf{0.759}} & {\color[HTML]{000000} 0.670} & {\color[HTML]{FE0000} \textbf{0.777}} \\
\multicolumn{1}{c|}{{\color[HTML]{000000} }} & {\color[HTML]{000000} RMSE} & {\color[HTML]{000000} 1.208} & {\color[HTML]{000000} 1.083} & {\color[HTML]{000000} 1.286} & {\color[HTML]{3166FF} \textbf{1.051}} & {\color[HTML]{000000} 1.139} & {\color[HTML]{FE0000} \textbf{1.010}} \\
\multicolumn{1}{c|}{\multirow{-4}{*}{{\color[HTML]{000000} MSU}}} & {\color[HTML]{000000} KRCC} & {\color[HTML]{000000} 0.498} & {\color[HTML]{000000} 0.555} & {\color[HTML]{000000} 0.423} & {\color[HTML]{3166FF} \textbf{0.568}} & {\color[HTML]{000000} 0.536} & {\color[HTML]{FE0000} \textbf{0.584}} \\ \hline
\multicolumn{1}{c|}{{\color[HTML]{000000} }} & {\color[HTML]{000000} PLCC} & {\color[HTML]{000000} 0.816} & {\color[HTML]{000000} 0.824} & {\color[HTML]{000000} 0.649} & {\color[HTML]{3166FF} \textbf{0.846}} & {\color[HTML]{000000} 0.816} & {\color[HTML]{FE0000} \textbf{0.854}} \\
\multicolumn{1}{c|}{{\color[HTML]{000000} }} & {\color[HTML]{000000} SRCC} & {\color[HTML]{000000} 0.812} & {\color[HTML]{000000} 0.821} & {\color[HTML]{000000} 0.612} & {\color[HTML]{3166FF} \textbf{0.843}} & {\color[HTML]{000000} 0.811} & {\color[HTML]{FE0000} \textbf{0.852}} \\
\multicolumn{1}{c|}{{\color[HTML]{000000} }} & {\color[HTML]{000000} RMSE} & {\color[HTML]{000000} 0.377} & {\color[HTML]{000000} 0.371} & {\color[HTML]{000000} 0.498} & {\color[HTML]{3166FF} \textbf{0.349}} & {\color[HTML]{000000} 0.378} & {\color[HTML]{FE0000} \textbf{0.340}} \\
\multicolumn{1}{c|}{\multirow{-4}{*}{{\color[HTML]{000000} YouTubeUGC}}} & {\color[HTML]{000000} KRCC} & {\color[HTML]{000000} 0.617} & {\color[HTML]{000000} 0.627} & {\color[HTML]{000000} 0.438} & {\color[HTML]{3166FF} \textbf{0.651}} & {\color[HTML]{000000} 0.618} & {\color[HTML]{FE0000} \textbf{0.662}} \\ \hline
\multicolumn{1}{c|}{{\color[HTML]{000000} }} & {\color[HTML]{000000} PLCC} & {\color[HTML]{3166FF} \textbf{0.938}} & {\color[HTML]{000000} 0.935} & {\color[HTML]{000000} 0.698} & {\color[HTML]{000000} 0.930} & {\color[HTML]{000000} 0.921} & {\color[HTML]{FE0000} \textbf{0.945}} \\
\multicolumn{1}{c|}{{\color[HTML]{000000} }} & {\color[HTML]{000000} SRCC} & {\color[HTML]{3166FF} \textbf{0.935}} & {\color[HTML]{000000} 0.932} & {\color[HTML]{000000} 0.685} & {\color[HTML]{000000} 0.927} & {\color[HTML]{000000} 0.918} & {\color[HTML]{FE0000} \textbf{0.942}} \\
\multicolumn{1}{c|}{{\color[HTML]{000000} }} & {\color[HTML]{000000} RMSE} & {\color[HTML]{3166FF} \textbf{4.969}} & {\color[HTML]{000000} 5.114} & {\color[HTML]{000000} 10.091} & {\color[HTML]{000000} 5.269} & {\color[HTML]{000000} 5.484} & {\color[HTML]{FE0000} \textbf{4.706}} \\
\multicolumn{1}{c|}{\multirow{-4}{*}{{\color[HTML]{000000} LIVE-WC}}} & {\color[HTML]{000000} KRCC} & {\color[HTML]{3166FF} \textbf{0.780}} & {\color[HTML]{000000} 0.773} & {\color[HTML]{000000} 0.507} & {\color[HTML]{000000} 0.771} & {\color[HTML]{000000} 0.758} & {\color[HTML]{FE0000} \textbf{0.794}} \\ \hline

\multicolumn{1}{c|}{{\color[HTML]{000000} }} & {\color[HTML]{000000} PLCC} & {\color[HTML]{000000} 0.856} & {\color[HTML]{000000} 0.892} & {\color[HTML]{000000} 0.529} & {\color[HTML]{3166FF} \textbf{0.907}} & {\color[HTML]{000000} 0.901} & {\color[HTML]{FE0000} \textbf{0.914}} \\
\multicolumn{1}{c|}{{\color[HTML]{000000} }} & {\color[HTML]{000000} SRCC} & {\color[HTML]{000000} 0.844} & {\color[HTML]{000000} 0.887} & {\color[HTML]{000000} 0.482} & {\color[HTML]{3166FF} \textbf{0.899}} & {\color[HTML]{000000} 0.894} & {\color[HTML]{FE0000} \textbf{0.908}} \\
\multicolumn{1}{c|}{{\color[HTML]{000000} }} & {\color[HTML]{000000} RMSE} & {\color[HTML]{000000} 6.515} & {\color[HTML]{000000} 5.705} & {\color[HTML]{000000} 10.837} & {\color[HTML]{3166FF} \textbf{5.335}} & {\color[HTML]{000000} 5.538} & {\color[HTML]{FE0000} \textbf{5.169}} \\
\multicolumn{1}{c|}{\multirow{-4}{*}{{\color[HTML]{000000} LIVE-APV}}} & {\color[HTML]{000000} KRCC} & {\color[HTML]{000000} 0.666} & {\color[HTML]{000000} 0.715} & {\color[HTML]{000000} 0.339} & {\color[HTML]{3166FF} \textbf{0.731}} & {\color[HTML]{000000} 0.717} & {\color[HTML]{FE0000} \textbf{0.717}} \\ \hline
\end{tabular}%
}
\vspace{-15pt}
\end{table}

\subsection{Ablation Study}

To verify the effectiveness of each core module in Tao-QoE and quantify their contributions to overall performance, we designed systematic ablation experiments. We used the complete Tao-QoE model as the baseline. This model is noted as ALL. It includes three modules: the semantic feature extraction module S, the optical flow motion feature extraction module FM, and the multi-scale feature fusion module F. Note that the feature regression module, which maps the fused features to the final QoE score, is a necessary component included in all model variants and is therefore omitted from the shorthand notation for brevity. Using the control variable method, we gradually removed key modules. We conducted a comprehensive evaluation across 15 databases as shown in TABLE~\ref{tab:ablation of QoE databases} and TABLE~\ref{tab:ablation of VQA databases}. 

\subsubsection{The Fundamental Role of the Semantic Feature Extraction Module (S)}
Semantic information forms the foundation for perceiving video content quality. Experimental results are shown in TABLE~\ref{tab:ablation of QoE databases} and TABLE~\ref{tab:ablation of VQA databases}. Removing the S module (FM) causes the most significant performance drops across all databases. In QoE tasks, the average SRCC decrease $\Delta$ is 0.201. In VQA tasks, the average $\Delta$ is 0.223. For example, on the WaterlooSQoE-III database, performance falls sharply from 0.890 to 0.511. On the LIVE-APV database, it drops from 0.908 to 0.482. These results strongly prove that the high-level semantic features extracted by the S module contribute the most to the model's high-quality predictions. The module captures content-related quality cues through a deep convolutional network. These cues include texture details, encoding distortion, and scene complexity.
These cues form the basic framework for quality assessment.
Without this module, the model loses its basic ability to understand the inherent quality attributes of video.

\subsubsection{The Optical Flow Motion module (FM) is specially designed for modeling temporal distortions}

The FM module explicitly captures motion information between video frames. Experimental results are shown in TABLE~\ref{tab:ablation of QoE databases} and TABLE~\ref{tab:ablation of VQA databases}. This ability is vital for detecting temporal issues like stalling event and jitter. We evaluate its contribution by removing the FM module and its input branch (comparing models ALL and S+F). On QoE databases featuring stalling events, removing FM causes a greater performance decline. The average decrease in SRCC ($\Delta$) is 0.043. On traditional VQA databases, the average decrease is only 0.029. Specifically, on the LIVE-NFLX-II database which is dominated by stalling distortions, SRCC drops by 0.168. In contrast, on the LIVE-Qualcomm database which is dominated by spatial distortions, SRCC drops by only 0.031. These results show that the FM module is not a general feature extractor. Instead, it serves as an efficient detector specifically for temporal motion distortions. The module processes pre-computed optical flow maps. It transforms pixel-level motion vectors into high-level motion patterns. This enables the model to accurately identify and evaluate playback smoothness issues caused by network or rendering problems. Comparison with methods that directly extract motion features from raw frames also shows a clear advantage. Explicit optical flow calculation provides a clearer and more easily separable motion representation. Therefore, it leads to better performance (as seen when comparing models ALL and S+F+M on most databases).

\subsubsection{The Multi-scale Feature Fusion module (F) is responsible for perceiving and integrating quality dynamics.}

This module performs differencing and fusion of features across time. Its goal is to capture sharp changes and dynamic variations in quality over time. An example is a quality switch in a video stream. For the database WaterlooSQoE-II, which specifically focuses on quality switches, removing module F caused the SRCC to drop by 0.043. However, we also tested on a typical VQA database without such deliberate switching. There, the performance drop was much smaller, with a $\Delta$ of 0.0054. This confirms that module F is a key component for perceiving non-stationary quality changes over time. The module uses multi-scale temporal differencing operations. This enhances the model's sensitivity to quality differences between adjacent segments. Therefore, it can effectively respond to changes in quality level. These changes are often caused by mechanisms like adaptive bitrate adjustment. This explains why the full model (ALL) consistently outperforms the S+FM model. The S+FM model only uses simple feature concatenation.

In summary, the systematic ablation study empirically validates the necessity and specific functional roles of each core module in Tao-QoE. The semantic feature module serves as the foundation of the system. The optical flow motion module acts as a specialized tool for temporal distortions. The multi-scale fusion module is essential for perceiving quality dynamics. The effective integration of these three modules enables Tao-QoE to achieve comprehensive and outstanding performance in video quality assessment.

\begin{table}[ht]
\centering
\caption{Computation time for each subnetwork on a single 1080×1920 video with 300 frames. Among them, for Tao-QoE (ALL), "Extract Motion Feature" denotes extracting motion features from the inter-frame optical flow sequence. For Tao-QoE (S+F+M), it denotes extracting motion features from the video frame sequence.}
\label{tab:subnetwork-computation-time}
\begin{tabular}{c|cc|cc}
\hline
\multicolumn{1}{c|}{\multirow{2}{*}{{\color[HTML]{000000} Subnetwork}}} & \multicolumn{2}{c}{Tao-QoE(ALL)} & \multicolumn{2}{c}{Tao-QoE(S+F+M)} \\ 
\cline{2-3} \cline{4-5}
        & Time(s) & GFLOPS & Time(s) & GFLOPS \\
\hline
Read Video             & 1.40   & 0.92   & 1.45   & 0.92   \\
Extract Optical Flow   & 20.88  & 1048.96& —      & —      \\
Extract Motion Feature & 1.24   & 101.63 & 1.71   & 101.63 \\
Extract Semantic Feature & 3.22 & 56.84  & 3.10   & 56.84  \\
Fusion and Regression  & 0.14   & 0.55   & 0.14   & 0.55   \\\hline
TOTAL         & 26.88 & 1208.90 & 6.40 & 159.94 \\
\hline
\end{tabular}
\vspace{-15pt}
\end{table}

\begin{table}[h]
\centering
\caption{Computation time for different models on a single 1080×1920 video with 300 frames}
\label{tab:Computation time for different models}
\begin{tabular}{c|ccc}
\hline
Models         & Times(s) & GFLOPS  & FPS   \\ \hline
ASPECT         & 8.05     & 61.38   & 37.27 \\
TLVQM          & 358.00   & 170.40  & 0.84  \\
VSFA           & 69.39    & 995.32  & 6.49  \\
SimpleVQA      & 10.33    & 119.49  & 29.04 \\
FastVQA        & 0.05     & 279.00  & 6000  \\ \hline
Tao-QoE(S+F+M) & 6.40     & 159.94  & 46.88 \\
Tao-QoE(ALL)   & 26.88    & 1208.90 & 11.16 \\ \hline
\end{tabular}
\end{table}

\subsection{Computational Complexity}
To measure the actual computation time, we used a test video. This video has 300 frames and is in 1080P resolution. Among the compared algorithms, TLVQM was implemented on the MATLAB platform. The other models (using GPU) were implemented in Python. All experiments were conducted on the same hardware setup. The hardware configuration is as follows: an Intel(R) Xeon(R) Silver 4214R 2.40GHz processor, 256GB of memory, and an NVIDIA GeForce RTX 3090 graphics card. This study focuses on live streaming applications. Therefore, we analyzed the computational cost and giga floating-point operations (GFLOPS) of the compared models. 

Computation times for each subnetwork are shown in Table ~\ref{tab:subnetwork-computation-time}. For the Tao-QoE (ALL) model, most of its computation time is spent on optical flow extraction. This step takes 77.68\% of the total computation time. Although using optical flow can significantly improve the model's prediction performance, it also reduces its real-time processing capability. In comparison, the Tao-QoE (S+F+M) model can process 46.88 frames per second (FPS). This meets the basic requirement for real-time prediction. The prediction performance of Tao-QoE (S+F+M) is slightly lower than that of Tao-QoE (ALL). Still, it is clearly better than the other compared models in this study. Therefore, in real-world applications, if high accuracy is needed and real-time performance is not strictly required, Tao-QoE (ALL) is recommended. On the other hand, if the system has strict real-time requirements, Tao-QoE (S+F+M) can be chosen. This helps achieve a better balance between performance and efficiency.

Computation times for other models are shown in Table ~\ref{tab:Computation time for different models}. In scenarios that require the highest real-time performance, FastVQA is the best choice. In scenarios that need to balance performance and real-time capability, Tao-QoE(S+F+M) is an ideal solution. Tao-QoE(ALL) provides stronger performance but sacrifices real-time speed. It is suitable for situations where high accuracy is required, and latency tolerance is relatively lenient. The current version of the Tao-QoE (ALL) model cannot process videos in real time. However, its architecture design has the potential to achieve real-time performance through further optimization. ~\cite{tu2021ugc} points out that many frame-based VQA algorithms can greatly reduce computation. This is especially true for deep learning models. They can do this by lowering the input frame rate. Experiments show that this down-sampling has little effect on the final quality prediction scores. Scores such as SROCC and PLCC remain almost unchanged. We can achieve real-time performance through several methods. For example, we can downsample the video. We can also resize the video frames. Alternatively, we can choose a lightweight optical flow model.

\section{Conclusion} 

In this study, we built a database for large-scale live streaming scenarios, called the TaoLive QoE Database. The database selected 42 high-quality videos as source videos. We changed the CRF parameter of the videos to simulate compression distortion. We also changed the PTS of the video frames to add distortions such as stalling events and accelerated playing. It is worth noting that uneven PTS distribution caused by network issues is the underlying reason for distortions like stalling events in live streaming. At the same time, we conducted subjective experiments to collect QoE scores for these videos. We randomly divided the subjects into two groups. We then calculated the correlation between the two groups. This proved the validity of the data. Additionally, we proposed a QoE model named Tao-QoE. This model extracts features from both semantic and motion aspects. It then evaluates the QoE of videos through feature fusion. We applied optical flow to extract motion features. Extensive experiments show that using optical flow motion features can significantly improve the model's prediction performance.

Our work also has several limitations that warrant further investigation. First, the video duration in our database is relatively short (10–22 seconds), which primarily supports retrospective QoE assessment. Extending the duration is nontrivial, as longer videos would significantly increase the complexity of subjective experiments and introduce greater variability in user attention and memory effects. Second, we recruited 20 participants following ITU-T P910 standards, with post-screening ensuring reliable raters. The subjective data quality is supported by narrow confidence intervals (57.4\% of videos have width $<$ 0.3) and high cross-group correlation (average SROCC $>$ 0.94). While this sample size is adequate for statistical reliability, we acknowledge that a larger and more diverse cohort would further enhance generalizability. Third, although Tao-QoE achieves strong prediction performance, its real-time capability is constrained by the computational cost of optical flow extraction. Achieving an optimal trade-off between accuracy and efficiency remains an open challenge that may require model compression, input downsampling, or lightweight optical flow alternatives. Fourth, other live streaming-specific distortions—such as macroblocking, blurring due to bitrate starvation, and ABR switching artifacts—are not covered in the current study. These distortions interact with stalling and encoding effects in complex ways, and their integration into a unified QoE framework poses both modeling and data acquisition challenges.

Addressing these limitations will guide our future research. Specifically, we plan to enrich the database by incorporating longer videos with a wider variety of distortions and network-simulated impairments. We will also expand the subjective test scale and explore continuous QoE prediction. For the model, we intend to investigate lightweight architectures and efficient motion representations to better balance performance and real-time deployability. More broadly, extending the current framework to encompass a richer set of live streaming distortions remains a key long-term objective.

\bibliography{IEEE-Transactions-LaTeX2e-templates-and-instructions/ref} 
\bibliographystyle{IEEEtran} 


\vspace{-12mm}

\begin{IEEEbiography}[{\includegraphics[width=1in,height=1.25in,clip,keepaspectratio]{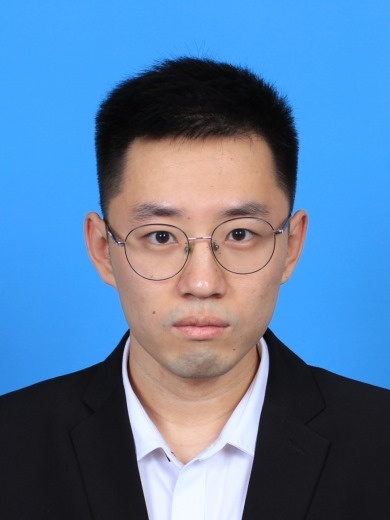}}]{Zehao Zhu}
received the B.E. degree in electronic information engineering from Jilin University in 2018 and the M.E. degree in information and communication engineering from Shanghai Jiao Tong University in 2021. He is currently pursuing the Ph.D. degree in electronic engineering with Shanghai Jiao Tong University, Shanghai, China. His research interests include streaming media quality of experience assessment and video quality assessment.
\end{IEEEbiography}

\vspace{-12mm} 

\begin{IEEEbiography}
[{\includegraphics[width=1in,height=1.25in,clip,keepaspectratio]{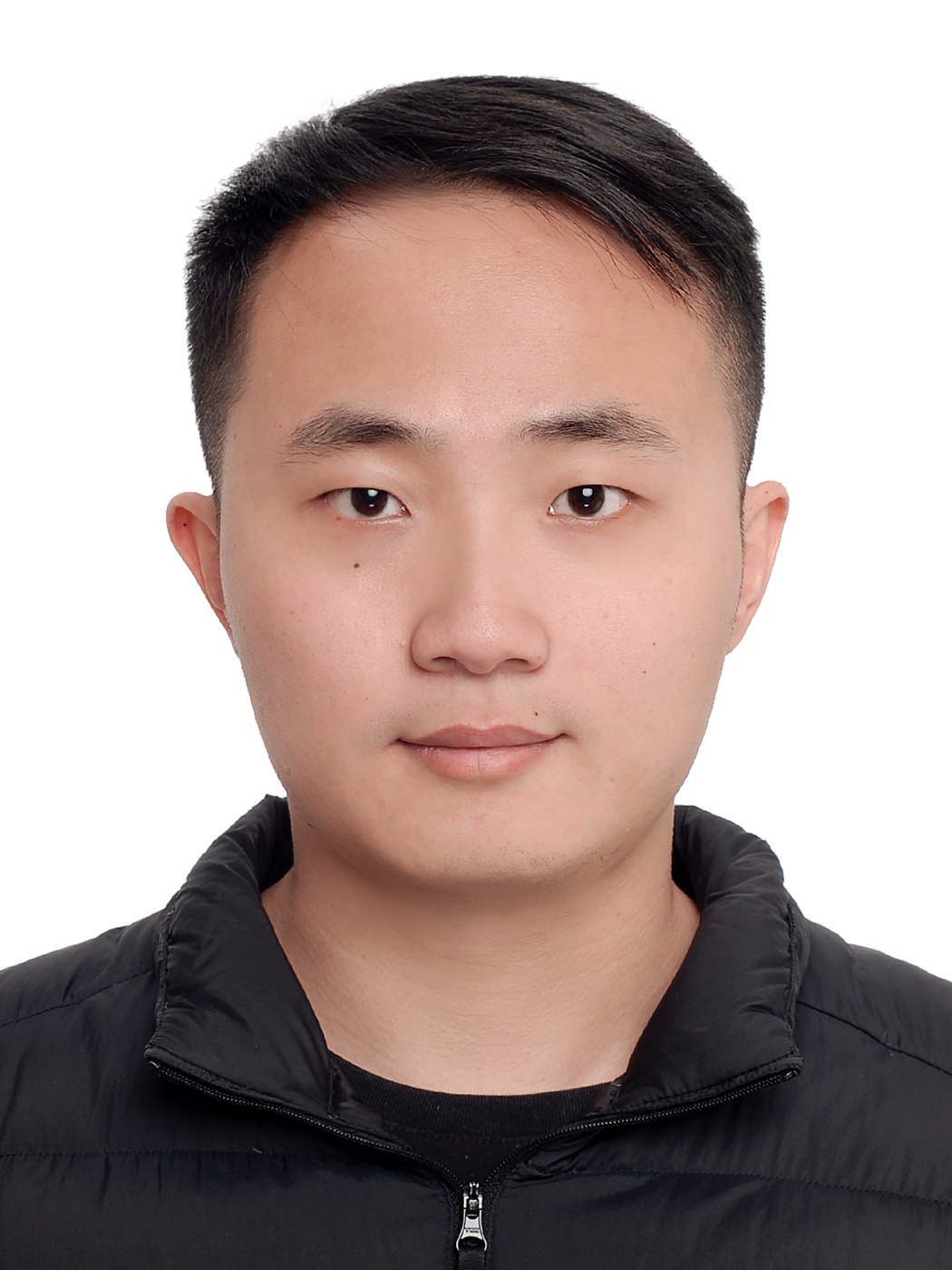}}]{Wei Sun}
received the B.E. degree from the East China University of Science and Technology, Shanghai, China, in 2016, and the Ph.D. degree from Shanghai Jiao Tong University, Shanghai, China, in 2023. He is currently a Post-Doctoral Fellow with Shanghai Jiao Tong University. His research interests include image quality assessment, perceptual signal processing and mobile video processing.
\end{IEEEbiography}

\vspace{-12mm} 

\begin{IEEEbiography}
[{\includegraphics[width=1in,height=1.25in,clip,keepaspectratio]{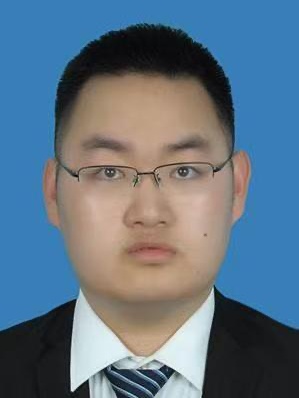}}]{Jun Jia}
received the B.S. degree in computer science and technology from Hunan University, Changsha, China, in 2018. He is currently pursuing the Ph.D. degree in electronic engineering with Shanghai Jiao Tong University, Shanghai, China.
His current research interests include computer vision and image processing.
\end{IEEEbiography}

\vspace{-12mm} 

\begin{IEEEbiography}
[{\includegraphics[width=1in,height=1.25in,clip,keepaspectratio]{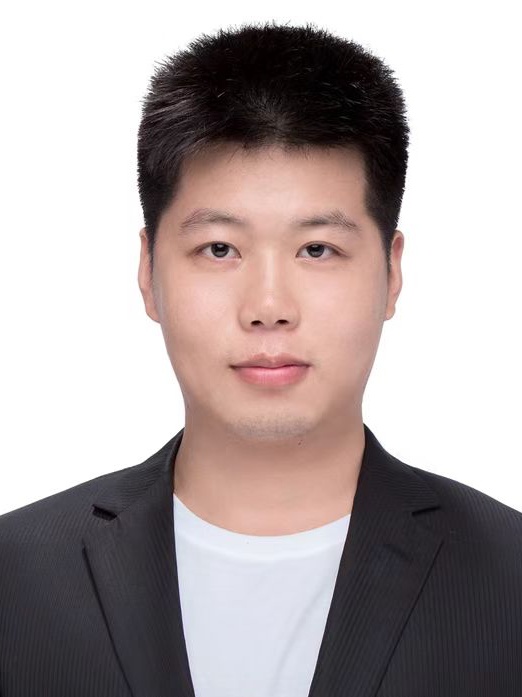}}]{Wei Wu}
received his Ph.D. in computer application technology from Wuhan University, Wuhan, China, in 2021. He joined Donghai Laboratory in 2023. Before joining Donghai Laboratory, he was a Senior Engineer with Alibaba Group, Hangzhou, China, from 2021 to 2023. His research interests include image and video processing, computer vision, and multi-source heterogeneous data fusion.
\end{IEEEbiography}

\vspace{-12mm}

\begin{IEEEbiography}
[{\includegraphics[width=1in,height=1.25in,clip,keepaspectratio]{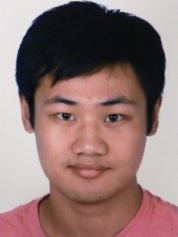}}]{Sibin Deng}
received the B.E. degree from Southeast University, Nanjing, China, in 2012, and the Ph.D. degree from the University of Science and Technology of China, Hefei, China, in 2017. He is currently an Engineer with Alibaba. His research interests include computer vision, image processing, and machine learning.
\end{IEEEbiography}

\vspace{-12mm} 

\begin{IEEEbiography}
[{\includegraphics[width=1in,height=1.25in,clip,keepaspectratio]{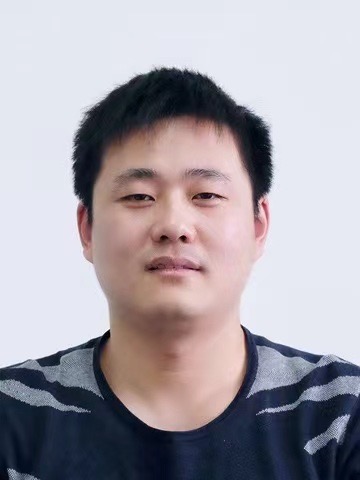}}]{Kai Li}
received the B.E. and PH.D. degrees from University of Science and Technology of China, Hefei, China, in 2004 and 2009. He is currently a Senior Expert of Algorithm worked for Zhejiang Tmall Technology Co., Ltd.
\end{IEEEbiography}

\vspace{-10mm} 

\begin{IEEEbiography}
[{\includegraphics[width=1in,height=1.25in,clip,keepaspectratio]{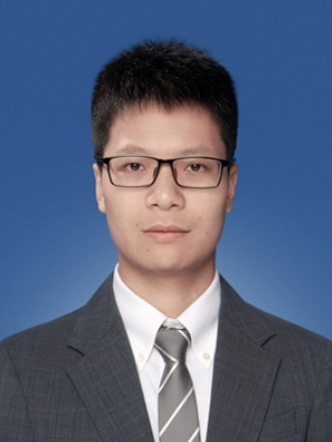}}]{Xiongkuo Min}
(Member, IEEE) received the B.E.
degree from Wuhan University, Wuhan, China, in
2013, and the Ph.D. degree from Shanghai Jiao
Tong University, Shanghai, China, in 2018. He is
currently a tenure-track Associate Professor with
the Institute of Image Communication and Network
Engineering, Shanghai Jiao Tong University. His
research interests include image/video/audio quality
assessment, quality of experience, visual attention
modeling, extended reality, and multimodal signal
processing. He received the Best Paper Runner-Up
Award of IEEE TRANSACTIONS ON MULTIMEDIA in 2021, the Best
Student Paper Award of the IEEE International Conference on Multimedia
and Expo (ICME) in 2016, and the Excellent Ph.D. Thesis Award from the
Chinese Institute of Electronics (CIE) in 2020.
\end{IEEEbiography}

\vspace{-12mm}

\begin{IEEEbiography}
[{\includegraphics[width=1in,height=1.25in,clip,keepaspectratio]{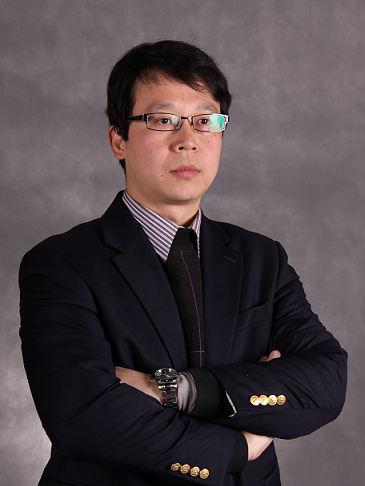}}]{Jia Wang}
received the B.Sc. degree in electronic engineering, the M.S. degree in pattern recognition and intelligence control, and the Ph.D. degree in electronic engineering from Shanghai Jiao Tong University, China, in 1997, 1999, and 2002, respectively. He is currently a Professor with the Department of Electronic Engineering, Shanghai Jiao Tong University, and also a member of the Shanghai Key Laboratory of Digital Media Processing and Transmission. His research interests include multiuser information theory and mathematics in artificial intelligence.
\end{IEEEbiography}

\vspace{-12mm}

\begin{IEEEbiography}
[{\includegraphics[width=1in,height=1.25in,clip,keepaspectratio]{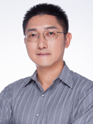}}]{Ying Chen}
(IEEE M’05 - SM’11) received a B.S. in Applied Mathematics and an M.S. in Electrical Engineering \& Computer Science, both from Peking University, in 2001 and 2004 respectively. He received his PhD in Computing and Electrical Engineering from Tampere University of Technology (TUT), Finland, in 2010.
Dr. Chen joined Alibaba Group, in 2018 as a Senior Director. Before joining Alibaba. His earlier working experiences include Principal Engineer in Qualcomm Incorporated, San Diego, CA, USA from 2009 to 2018, Researcher in TUT and Nokia Research Center, Finland from 2006 to 2009 and Research Engineer in Thomson Corporate Research, Beijing, from 2004 to 2006. Dr. Chen is currently leading the Audiovisual Technology Group in Taobao, Alibaba, supporting end-to-end multimedia features and applications within Taobao. Dr. Chen has been focusing on algorithm development and commercialization of multimedia technologies. Dr. Chen contributed to three generations of video coding standards, H.264/AVC, H.265/HEVC, and H.266/VVC (earlier stage) and video file format and transport standards, with various technical contribution documents (500+). Dr. Chen has served as an editor and a software coordinator for H.264/AVC and H.265/HEVC (both for Multiview and 3D Video extensions). Dr. Chen’s research areas include video coding, image/video restoration and enhancement, image/video quality assessment and video transmission. Dr. Chen has authored or co-authored 80+ academic papers and over 250 granted US patents in the fields of image/video processing and coding, multimedia transmission and computer vision. His publications have been cited for more than 20000 times. Dr. Chen has served as an associate editor for IEEE Transactions on CSVT.
\end{IEEEbiography}

\vspace{-12mm}

\begin{IEEEbiography}
[{\includegraphics[width=1in,height=1.25in,clip,keepaspectratio]{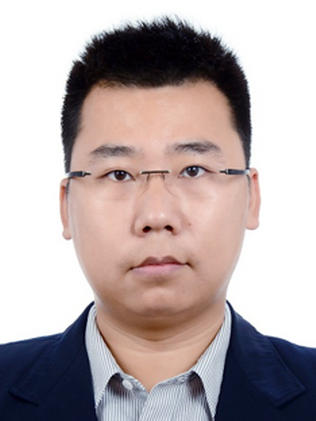}}]{Guangtao Zhai }
 (Fellow, IEEE) received the B.E. and M.E. degrees from Shandong University, Shandong, China, in 2001 and 2004, respectively, and the Ph.D. degree from Shanghai Jiao Tong University, Shanghai, China, in 2009. From 2008 to 2009, he was a Visiting Student with the Department of Electrical and Computer Engineering, McMaster University, Hamilton, ON, Canada, where he was a Post-Doctoral Fellow, from 2010 to 2012. From 2012 to 2013, he was a Humboldt Research Fellow with the Institute of Multimedia Communication and Signal Processing, Friedrich-Alexander-University of Erlangen–Nüremberg, Germany. He is currently a Research Professor with the Institute of Image Communication and Information Processing, Shanghai Jiao Tong University. His research interests include multimedia signal processing and perceptual signal processing. He received the Award of National Excellent Ph.D. Thesis from the Ministry of Education of China in 2012.
\end{IEEEbiography}

\end{document}